\documentclass[sigconf]{acmart}
\AtBeginDocument{%
  }

\copyrightyear{2026}
\acmYear{2026}
\setcopyright{cc}
\setcctype{by}
\acmConference[CHI '26]{Proceedings of the 2026 CHI Conference on Human Factors in Computing Systems}{April 13--17, 2026}{Barcelona, Spain}
\acmBooktitle{Proceedings of the 2026 CHI Conference on Human Factors in Computing Systems (CHI '26), April 13--17, 2026, Barcelona, Spain}
\acmPrice{}
\acmDOI{10.1145/3772318.3790932}
\acmISBN{979-8-4007-2278-3/2026/04}

\usepackage{multirow}
\usepackage[table,x11names]{xcolor}
\usepackage[flushleft]{threeparttable}
\usepackage{enumitem}
\usepackage{amsmath}
\usepackage{algorithm}
\usepackage{algpseudocode}

\colorlet{shadecolor}{lightgray!30}

\definecolor{red}{RGB}{0, 0, 0}
\definecolor{todo}{RGB}{0, 0, 0}
\definecolor{blue}{RGB}{0,0,0}

\begin{document}

\title{WatchHand: Enabling Continuous Hand Pose Tracking On Off-the-Shelf Smartwatches}


\author{Jiwan Kim}
\affiliation{%
  \institution{KAIST}
  \city{Daejeon}
  \country{Republic of Korea}}
  \email{jiwankim@kaist.ac.kr}
  \authornote{Contributed equally to this paper}

\author{Chi-Jung Lee}
\affiliation{%
 \institution{Cornell University}
 \city{Ithaca}
 \state{New York}
 \country{USA}}
 \email{cl2358@cornell.edu}
 \authornotemark[1]

\author{Hohurn Jung}
\affiliation{%
  \institution{KAIST}
  \city{Daejeon}
  \country{Republic of Korea}}
\email{cllocker@kaist.ac.kr}

\author{Tianhong Catherine Yu}
\affiliation{%
 \institution{Cornell University}
 \city{Ithaca}
 \state{New York}
 \country{USA}}
 \email{ty274@cornell.edu}

\author{Ruidong Zhang}
\affiliation{%
 \institution{Cornell University}
 \city{Ithaca}
 \state{New York}
 \country{USA}}
 \email{rz379@cornell.edu}

\author{Ian Oakley}
\affiliation{%
  \institution{KAIST}
  \city{Daejeon}
  \country{Republic of Korea}}
\email{ian.r.oakley@gmail.com}
\authornote{Co-corresponding authors}

\author{Cheng Zhang}
\affiliation{%
  \institution{Cornell University}
  \city{Ithaca}
  \state{New York}
  \country{USA}}
  \email{chengzhang@cornell.edu}
  \authornotemark[2]
\renewcommand{\shortauthors}{Kim, Lee, et al.}

\begin{abstract}
Tracking hand poses on wrist-wearables enables rich, expressive interactions, yet remains unavailable on commercial smartwatches, as prior implementations rely on external sensors or custom hardware, limiting their real-world applicability. To address this, we present WatchHand, the first continuous 3D hand pose tracking system implemented on off-the-shelf smartwatches using only their built-in speaker and microphone. WatchHand emits inaudible frequency-modulated continuous waves and captures their reflections from the hand. These acoustic signals are processed by a deep-learning model that estimates 3D hand poses for 20 finger joints. We evaluate WatchHand across diverse real-world conditions---multiple smartwatch models, wearing-hands, body postures, noise conditions, pose-variation protocols---and achieve a mean per-joint position error of 7.87 mm in cross-session tests with device remounting. \textcolor{red}{Although performance drops for unseen users or gestures, the model adapts effectively with lightweight fine-tuning on small amounts of data.} Overall, WatchHand lowers the barrier to smartwatch-based hand tracking by eliminating additional hardware while enabling robust, \textcolor{red}{always-available} interactions on millions of existing devices.

\end{abstract}



\begin{teaserfigure}
  \includegraphics[width=\textwidth]{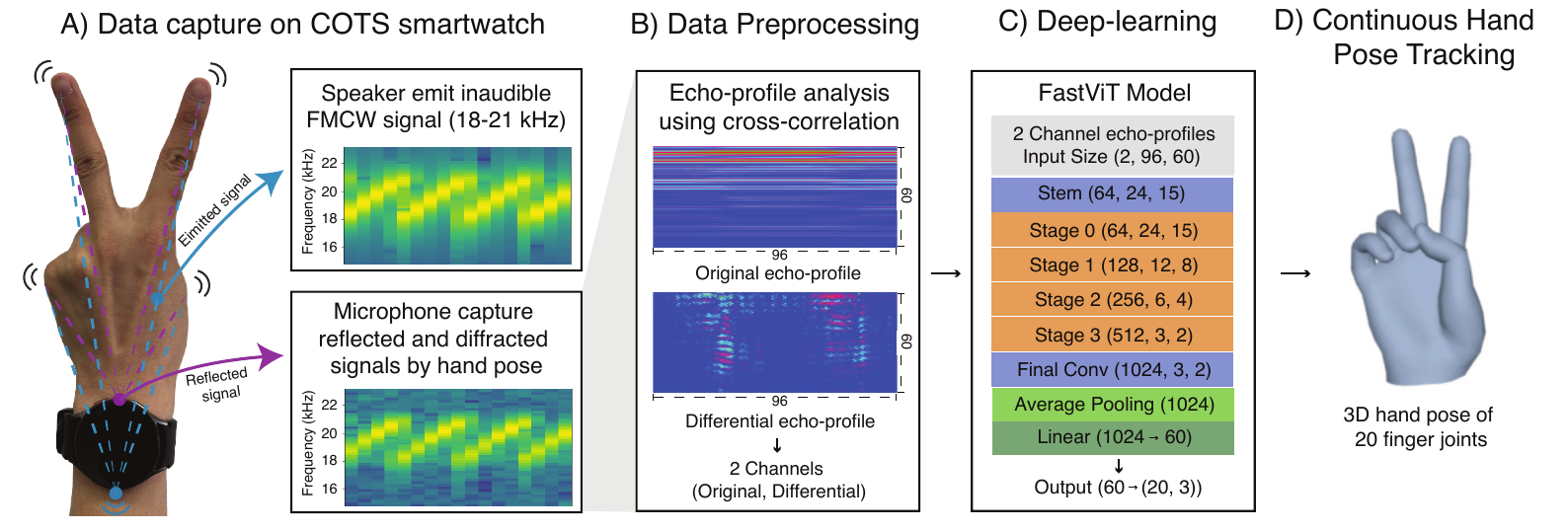}
  \caption{Overview of WatchHand: (A) Commercial off-the-shelf smartwatch emits inaudible frequency-modulated continuous waves (FMCW) via its speaker while capturing reflected and diffracted signals from hand motion through its microphone. (B) The captured acoustic signals are preprocessed using cross-correlation to generate echo profiles---a spatiotemporal representation of signal reflections. (C) A deep-learning model then processes echo profiles to (D) continuously track 3D hand poses.}
  \Description{
  Figure 1 illustrates the overall architecture of WatchHand, a system for continuous 3D hand pose tracking using only built-in sensors on a commercial smartwatch. (A) The smartwatch emits inaudible FMCW signals (18–21 kHz) and captures their reflections and diffractions caused by varying hand poses. (B) These acoustic signals are transformed into echo profiles using cross-correlation, resulting in both raw and differential representations. (C) A two-channel FastViT-based model processes the spatiotemporal echo features to regress (D) the continuous 3D positions of 20 finger joints. This pipeline demonstrates the feasibility of achieving continuous hand tracking without additional hardware or external sensors.
  }
  \label{fig:teaser}
\end{teaserfigure}


\maketitle

\section{Introduction}
Hands are our fundamental tools for interacting with the surrounding world, expressing our physical intentions, and performing activities. However, despite millions of smartwatches on the market, none of these commercial devices can continuously track hand poses, a key technique to understanding user intentions and behaviors, such as supporting context-aware applications or behavioral communication and expression (e.g., finger-spelling~\cite{Pugeault11Spelling, lim2025spellring, kaifosh2025generic}). Recognizing the importance of tracking hand poses for interaction, researchers have explored various sensing methods using a wrist-mounted form factor, including cameras~\cite{12digitsKim, Devrio22DiscoBand, FingerIO16Nandakumar}, radio frequency (RF) antennas~\cite{Kim22EtherPose}, electromyography (EMG)~\cite{Liu21WRHand}, radar~\cite{RadarHand24Hajika}, active sonar~\cite{Lee24_echoWrist, SonarID22Kim}, and motion sensing~\cite{Serendipity16Wen, GestureCustomization22Xu}. However, all these prior works either focus solely on recognizing discrete hand gestures (rather than continuously tracking hand poses)~\cite{Serendipity16Wen, GestureCustomization22Xu, SonarID22Kim} or require additional hardware~\cite{12digitsKim, Devrio22DiscoBand, Hu20FingerTrak, Kim22EtherPose, Kyu24EITPose, Liu21WRHand, Lee24_echoWrist} that is not available on millions of existing commercial off-the-shelf (COTS) smartwatches. As a result, continuous hand pose tracking remains only available on customized hardware or in research prototypes. We believe that enabling continuous hand pose tracking on COTS smartwatches without requiring additional hardware would significantly advance research in this area by immediately unlocking a wide range of downstream applications that could substantially improve end-user interaction experience. 

The primary challenge of continuous hand pose tracking lies in capturing detailed information about all finger joint movements and hand dynamics. Recent work by Lee et al.~\cite{Lee24_echoWrist} demonstrated the feasibility of using two speaker-microphone pairs to form micro-sonars on a customized wristband, enabling ultrasound-based tracking of finger movements. Inspired by this, we observed that modern smartwatches typically contain at least one speaker and one microphone, which could form a single micro-sonar. We thus hypothesize that \textit{repurposing the built-in sensors can provide sufficient information to continuously track finger joint movements and hand poses on COTS smartwatches}. In this paper, we aim to evaluate this hypothesis by addressing several key challenges that must be tackled to adapt a prior proof-of-concept continuous hand pose tracking sonar implementation on bespoke hardware~\cite{Lee24_echoWrist} into a practical and deployable solution for existing COTS smartwatches. 

\begin{itemize} [labelindent=\parindent,itemindent=0pt,leftmargin=*]
    \item \textbf{Hardware configuration}: Commercial smartwatches typically include only a single speaker-microphone pair, often placed on opposite sides of the case (see Figure~\ref{fig:watches}). It remains an open question whether a single speaker-microphone pair on COTS smartwatches can capture sufficient reflections for fine-grained hand-pose estimation. Moreover, a practical sensing system needs to support different smartwatch models, varying in microphone and speaker quality, placement, and design.
    
    \item \textbf{Robustness to device re-wearing}:  When users remove and re-wear the device, small shifts in placement or orientation can alter sensor data patterns, leading to variability in tracking performance~\cite{Hu20FingerTrak, Devrio22DiscoBand, Kyu24EITPose}. A practical sensing system should provide reliable hand pose tracking performance even after the watch is remounted. 

    \item \textbf{\textcolor{blue}{Real-world} variability}: A practical sensing system must also remain robust under different wearing habits, postures, \textcolor{blue}{noises}, and motion patterns. 
    \begin{itemize} 
        \item Accounting for different watch-wearing hand preferences (e.g., left/right hand), which reverses the spatial orientation of the speaker and microphone.
        \item Supporting varied body postures, such as arm-raised to view the screen or arm-resting naturally beside the body, which may affect signal reflections and thus hand pose tracking performance. 
        \item \textcolor{blue}{Maintaining robustness under challenging conditions such as environmental (e.g., loud music or nearby human movement) or user-induced noises (e.g., walking conditions or altered hand poses).}
        \item Covering dynamic hand pose variations, including both pose-neutral-pose (i.e., returning to a fixed neutral pose between pose changes) and direct pose-to-pose transitions, as well as varying finger movement speeds.
    \end{itemize}

\end{itemize}

Despite many pioneering efforts, these real-world variations have not been fully addressed in prior work, yet they represent critical challenges towards building robust, generalizable, and scalable continuous 3D hand pose tracking systems that can make an immediate impact on millions of COTS smartwatches.

In this paper, we aim to tackle the above challenges and lower the barrier of continuous hand pose tracking on wrist-mounted devices by presenting WatchHand, the first continuous hand-pose tracking system for COTS smartwatches that relies solely on the built-in speaker and microphone. Our approach leverages the cross-correlation-based frequency-modulated continuous waves (C-FMCW) method~\cite{wang2018c}. The system emits inaudible FMCW signals (18–21 kHz) through its speaker \textcolor{blue}{into the air} and captures reflections caused by hand poses via its microphone. The signals are then analyzed with a cross-correlation-based method to generate spatiotemporal echo profiles. These profiles are fed into a deep learning model to continuously estimate 3D hand poses. To evaluate WatchHand, we adopted established study protocols from prior work that employed custom arrays of speakers and microphones~\cite{Lee24_echoWrist, Yu24_ringAPose}. To demonstrate WatchHand’s real-world applicability, we ran a first study (N = 24) that tested it on three recent COTS smartwatches from three different manufacturers (Samsung, Xiaomi, and Google), which were worn on either hand. We then conducted a follow-up second study by re-inviting 6 participants and investigating system performance across varying body postures. \textcolor{blue}{In the third study (N = 8), we assessed robustness under various noise conditions, including loud music, nearby human motion, walking, and altered hand poses.} Finally, we conducted a \textcolor{blue}{fourth} study (N = 8) to evaluate the system’s capability in tracking more dynamic hand pose variations. These studies collectively sought to establish performance in diverse real-world conditions. 

The results show WatchHand reliably tracks 3D hand poses, achieving a mean per-joint position error (MPJPE) of 7.8\textcolor{blue}{7} mm in a cross-session model \textcolor{red}{and 14.88 mm in a cross-user model} across 20 finger joints spanning different smartwatch models and watch-wearing hands. \textcolor{blue} {The system also remains robust under variations in arm and body postures and in noisy conditions with only a single session (2 minutes of data) of fine-tuning.} Furthermore, WatchHand covers dynamic hand pose variation (mean MPJPE of \textcolor{blue}{15.48} mm) in cross-session testing. To provide a comprehensive view of system performance, we further extended our analysis to within-session and cross-user protocols, enabling a direct comparison with prior works and highlighting WatchHand’s robustness across diverse evaluation settings. These findings confirm that WatchHand provides promising performance \textcolor{red}{even after the device is remounted, while revealing expected drops in cross-user performance or when exposed to dynamic gesture variations. Nevertheless, the system adapts effectively to different body postures and noise conditions,} highlighting its practicality and suitability for real-world hand-tracking applications. The dataset collected across four studies with 40 participants is publicly available at \textcolor{red}{\url{https://github.com/witlab-kaist/WatchHand}}. Consequently, by leveraging only pre-existing built-in hardware, WatchHand demonstrates the feasibility of enabling continuous hand-pose tracking on COTS smartwatches. Our contributions include:

\begin{itemize}

\item Significantly lowering the barrier for continuous hand pose tracking on wrist-mounted devices by being the first to enable and demonstrate a continuous 3D hand pose tracking system developed exclusively using the built-in speaker and microphone on off-the-shelf smartwatches, requiring no external sensors or hardware modifications.

\item The design and evaluation of deep learning models for hand pose estimation with multiple training strategies, including within-session, cross-session, and cross-user approaches to support broad comparisons with prior work.

\item \textcolor{blue}{Extensive} empirical evaluations using three widely used smartwatch brands (Samsung, Xiaomi, and Google). The promising performance exhibited across diverse scenarios (e.g., different watch models, wearing hands, body postures, \textcolor{blue}{noise conditions}, and hand pose variation protocols) demonstrates WatchHand's real-world feasibility. \textcolor{blue}{We released the dataset, comprising a total of 35.6 hours of hand pose variations captured from 40 participants across all studies, to support further research in this area.}

\item Discussion of key insights toward generalization across diverse hand poses, unseen users, and use contexts, along with practical considerations, such as power consumption, privacy, system deployment, and application designs that can support real-world adoption. 

\end{itemize}
\section{Related Work}

\begin{table*}[t!]
\begin{threeparttable}
\centering
\setlength{\tabcolsep}{12 pt}
\Description{Table 1 presents a comprehensive comparison of wrist-worn systems for continuous hand pose tracking, focusing on sensing modality, number of sensors, study protocol (Pose-N-Pose vs. Pose-to-Pose), and evaluation models (within-session, cross-session, and cross-user models). WatchHand is uniquely evaluated under both protocols, demonstrating its robustness. Under the Pose-N-Pose protocol, it achieves 6.06 mm within-session, 7.87 mm cross-session, and 14.88 mm cross-user error, significantly outperforming many systems that require bulky or external hardware. Under the more challenging Pose-to-Pose protocol, it maintains competitive accuracy with 12.68 mm within-session, 15.48 mm cross-session, and 22.60 mm cross-user error, illustrating generalizability under continuous motion. Unlike prior work, WatchHand uses only a single built-in speaker, mic, and IMU on commercial off-the-shelf (COTS) smartwatches—with no added hardware—making it far more practical for deployment. 
}
\caption{An overview of existing wrist-worn continuous hand pose tracking systems. The Pose-N-Pose protocol denotes returning to a neutral (default) pose between variations, while Pose to Pose signifies direct transitions between target hand poses. Cross-session evaluation refers to session-independent testing with device re-mounting~\cite{Hu20FingerTrak, Lee24_echoWrist} or across different days~\cite{Kyu24EITPose}. Errors are reported in MPJPE (mm), except for * denoting MPJAE ($^\circ$).}
        \begin{tabular}{c|c|c|c|c|c}
         \multirow{2}{*}{Related Work} & \multirow{2}{*}{\shortstack[2]{Sening Modality \\($\#$ of sensors)}} & \multirow{2}{*}{\shortstack[2]{Study Protocol \\ ($\#$ of Ref Poses)}} & \multirow{2}{*}{\shortstack[2]{Within Session\\ Error}} & \multirow{2}{*}{\shortstack[2]{Cross Session\\ Error}} & \multirow{2}{*}{\shortstack[2]{Cross User \\ Error}} \\ &&&&& \\
        \hline
        FingerTrak~\cite{Hu20FingerTrak} & Thermal cameras (4)  & Pose-N-Pose (19) & 12 mm & 27.2mm  & 87 mm \\
        WR-Hand~\cite{Liu21WRHand}  & EMG sensors (8), gyro (1) & Pose-N-Pose (12) & 25.7 mm & - & 31.3 mm \\
        EchoWrist~\cite{Lee24_echoWrist} & Speakers (2), mics (2) &  Pose-N-Pose (15) & - & 4.81 mm & 12.2 mm \\
        \rowcolor{shadecolor} & & Pose-N-Pose (15) & \textcolor{blue}{6.06} mm & 7.8\textcolor{blue}{7} mm & 1\textcolor{blue}{4.88} mm\\
        \rowcolor{shadecolor} \multirow{-2}{*}{\textbf{WatchHand}} & \multirow{-2}{*}{\shortstack[2]{COTS smartwatch's \\built-in speaker (1), mic (1)}} & Pose to Pose (15) & 1\textcolor{blue}{2.68} mm & 1\textcolor{blue}{5.48} mm & 2\textcolor{blue}{2.60} mm \\
        Digits~\cite{12digitsKim} & IR camera (1), IR laser (1) & Pose to Pose (10) & 2$^\circ$ - 9$^\circ$* & - & - \\
        DiscoBand~\cite{Devrio22DiscoBand} & Depth cameras (8)& Pose to Pose (10) & 11.69 mm & 17.87 mm & 19.98 mm \\
        EtherPose~\cite{Kim22EtherPose} & RF antennas (2) & Pose to Pose (11) & 11.57 mm & - & - \\
        EITPose~\cite{Kyu24EITPose} & EIT electrodes (8) & Pose to Pose (12) & 11.06 mm & 17.81 mm & 18.91 mm \\
    \end{tabular}
    \label{tab:rw}
    \end{threeparttable}
\end{table*}

\subsection{Active Acoustic Sensing}
\label{sec:rw-active-acoustic-sensing}
Active acoustic sensing has emerged as a promising technique for around-device interaction by analyzing reflected or diffracted inaudible sound waves emitted from speakers to detect user gestures or movements. Early work such as SoundWave~\cite{Gupta12SoundWave} leveraged Doppler shifts and a laptop's built-in speaker and microphone to recognize mid-air gestures. Similarly, FingerIO~\cite{FingerIO16Nandakumar} and LLAP~\cite{LLAP16Wang} explored orthogonal frequency-division multiplexing (OFDM) or multiple continuous wave signals on customized smartwatches or commercial smartphones equipped with two pairs of microphones for precise near-device finger tracking. Building on these initial demonstrations, researchers have implemented richer acoustic sensing paradigms on various wearable devices, such as wrist-bands~\cite{Lee24_echoWrist, lee2025grab, Mahmoodi25EchoForce}, smartwatches~\cite{kim2025cross, SonarID22Kim, Zhang24LipWatch}, rings~\cite{Yu24_ringAPose}, smart-glasses~\cite{Mahmud24ActSonic, Zhang23EchoSpeech, mahmud2023posesonic}, and earbuds~\cite{Suzuki24EarHover}. This research has sought to enable more expressive and context-aware interaction paradigms, such as discrete gesture recognition~\cite{Suzuki24EarHover, SonarID22Kim, lee2025grab}, continuous hand pose tracking~\cite{Lee24_echoWrist, Yu24_ringAPose}, silent speech recognition~\cite{Zhang24LipWatch, Zhang23EchoSpeech}, and activity recognition~\cite{Mahmud24ActSonic}. 

Collectively, these studies highlight the potential of active acoustic sensing with a simple hardware configuration (using only speakers and microphones) to realize diverse and expressive interaction techniques. However, while their achievements are impressive, their reliance on specialized, bespoke hardware designs---often requiring multiple carefully positioned speaker and microphone pairs~\cite{Lee24_echoWrist, Zhang23EchoSpeech, Mahmud24ActSonic, FingerIO16Nandakumar}---limits practical integration into commercial devices. We argue that work in this area needs to move beyond proof-of-concept prototypes to tackle the real-world deployment challenges, such as relying on single, suboptimally positioned speakers and microphones (e.g., distant, opposite-facing). WatchHand directly addresses these limitations, utilizing only built-in smartwatch hardware to tackle the challenging task of 3D hand pose tracking. This advances the state of the art by demonstrating how active sonar systems can be implemented on current commercial wearable devices.

\subsection{Hand Pose Input on Commercial Smartwatches}

Smartwatch hand pose input involves using actions of the watch-wearing hand to issue commands or respond to events. It is a compelling paradigm as it inherently enables single-handed use~\cite{WrisText18Gong, Serendipity16Wen, GestureCustomization22Xu, kaifosh2025generic}, and rich expressivity through the high dexterity of our hands~\cite{Lee24_echoWrist, Kim22EtherPose, RadarHand24Hajika, Liu21WRHand, Devrio22DiscoBand}. However, sensing these fine-grained actions remains challenging. Prior work has predominantly focused on leveraging motion signals captured by a watch's IMU. For example, Serendipity~\cite{Serendipity16Wen} proposed a discrete five-class hand-pose recognition system using watch motion data with an average 87\% F1-score. Xu et.al. ~\cite{GestureCustomization22Xu} further introduced a user-adaptive method for personalizing smartwatch-based pose input that enabled detection of a more substantial set of discrete poses (e.g., 12 to 16) with up to 92.1\% F1 score. Tsunoda et al.\cite{tsunoda2024thumb} detected 11 thumb-to-finger gestures with 90.2\% accuracy using the watch's accelerometer. 

While these results are promising, their reliability in real-world applications remains in doubt, and commercial realizations of these techniques have tended to focus on highly limited gesture sets composed of simple poses such as thumb-index pinches~\cite{ApplewatchAssitiveTouch} or making a fist~\cite{samsungMoreThan} or rely on integrating additional sensors (e.g., such as PPG~\cite{Zhang18FinDroidHR}). We argue that the full potential of hand pose input as an expressive modality is inadequately realized by current motion sensing implementations. To fully unlock richer and expressive interactions, we need new systems that explore alternative sensing channels available on commercial devices, such as the sonar used in WatchHand. 

\subsection{Continuous Hand Pose Tracking on Wrist Wearables}

Continuous hand pose tracking enables expressive, seamless, and context-aware interactions, making it a valuable modality for wearable computing. Existing approaches have explored various around-device sensing techniques integrated into wrist-worn wearables, such as camera~\cite{Devrio22DiscoBand, 12digitsKim, Hu20FingerTrak, wu20backhandpose}, radar~\cite{RadarHand24Hajika}, sound~\cite{Lee24_echoWrist}, RF antenna~\cite{Kim22EtherPose}, proximity~\cite{WrisText18Gong}, or electromyography (EMG)~\cite{Liu21WRHand} to track our hand movements. Camera-based methods provide high precision but face significant practical constraints. They generally require dedicated space for camera placements, restrict interaction area to the camera's fields of view, are sensitive to lighting changes, and require a high computational cost for real-time image processing. Moreover, camera usage in wearables raises privacy concerns, particularly in public settings~\cite{Lehman22Privacy, Roesner14Privacy}. Alternative sensing modalities have thus been proposed to mitigate these issues. EtherPose~\cite{Kim22EtherPose} tracks hand poses using wrist-worn RF antennas, while WR-Hand~\cite{Liu21WRHand} analyzes EMG signals captured from forearm muscles.

However, these solutions require external sensors or specialized hardware, which are cumbersome and impractical for daily use. For instance, EMG-based systems require electrodes in constant skin contact with tight-fitting wristbands, impacting user comfort~\cite{Liu21WRHand}. Similarly, antenna-based~\cite{Kim22EtherPose} and radar-based methods~\cite{RadarHand24Hajika} involve additional bulky components that preclude seamless integration. Moreover, these systems often require careful sensor positioning, as slight shifts in placement can distort sensor signals or substantially degrade tracking reliability~\cite{Hu20FingerTrak, Devrio22DiscoBand, Kyu24EITPose}, which is unrealistic in everyday use, where users frequently remove and re-wear devices. Recently, EchoWrist~\cite{Lee24_echoWrist} introduced a wristband-based active-sonar system using two carefully positioned pairs of speakers and microphones, achieving accurate hand pose estimation (MPJPE of 4.81 mm) and enabling hand-object interaction, even across device re-mounting sessions. However, its reliance on bespoke hardware configurations---requiring spatial separation across both sides of the wrists---makes it impractical for the existing commercial smartwatches. 

Table~\ref{tab:rw} provides an overview of wrist-worn continuous 3D hand pose tracking techniques explored in related works. Based on this analysis, to the best of our knowledge, there is no existing continuous hand pose estimation system that is a viable deployment candidate for current (and likely future) smartwatch form factors. We propose WatchHand as the first solution that achieves continuous hand tracking on unmodified commercial smartwatches.
\section{System Design and Implementation}
\label{System}

\begin{figure*}[t]
\centering
   \includegraphics[width=\textwidth]{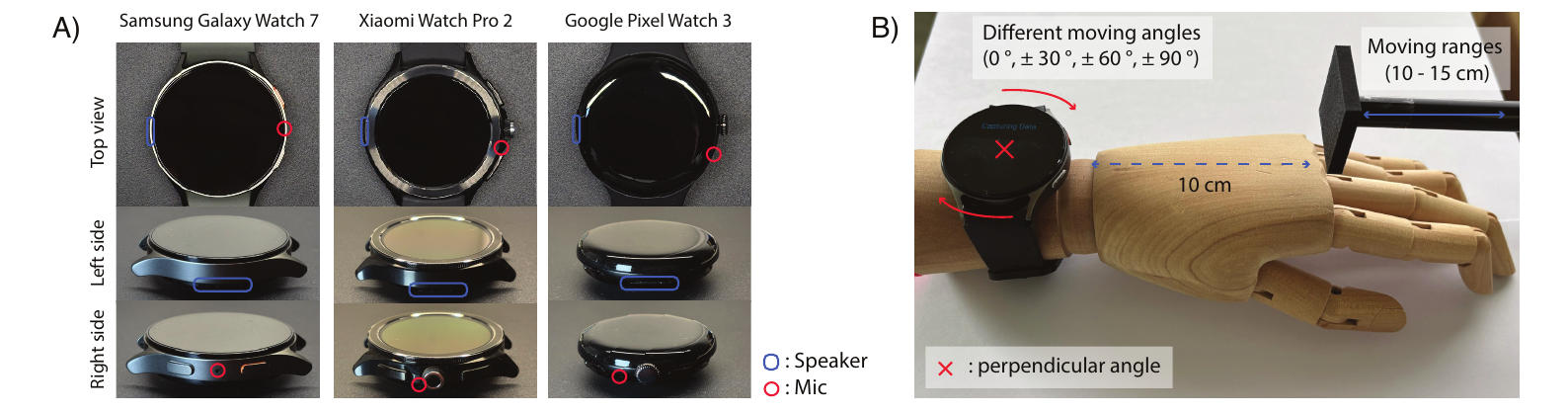}
   \hfil
\caption{(A) Three different COTS smartwatches we evaluated and the physical locations of their built-in speaker and microphone, and (B) system evaluation setup: A smartwatch is worn on a prop hand while a stepper motor–driven linear stage moves a flat plate back and forth within typical finger movement ranges (10–15 cm). To evaluate different angles, the prop hand was rotated to various orientations (e.g., 0°, ±30°, ±60°, ±90°, and perpendicular) relative to the direction of plate motion.}
\Description{Figure 2-A shows the physical locations of the built-in speaker and microphone on three commercial off-the-shelf (COTS) smartwatches: Samsung Galaxy Watch 7, Xiaomi Watch Pro 2, and Google Pixel Watch 3. The speaker (blue outline) is consistently located on the left side across all models, while the microphone (red outline) varies slightly in position on the right side. These placements are critical to understanding how the system captures acoustic signals for active sensing. Figure 2-B illustrates the system evaluation setup. A smartwatch is mounted on a prop hand while a stepper motor–driven linear stage moves a flat plate back and forth within typical finger movement ranges (10–15 cm). To evaluate directional sensitivity, the prop hand is rotated to various angles (e.g., 0°, ±30°, ±60°, ±90°, and perpendicular) relative to the direction of plate motion. This setup simulates realistic 3D hand interaction scenarios for assessing echo profile variability.}
\label{fig:watches}
\end{figure*}

\subsection{Design Consideration and Principle}
The main research question this paper aims to answer is: \textit{Can we enable an off-the-shelf smartwatch to continuously track hand poses?} This is a highly challenging task, as it requires accurately estimating the 3D positions of 20 finger joints using only the limited sensing utility of COTS smartwatches. To address this problem, we examine the most common sensors embedded into modern commercial smartwatches, including the microphone, speaker, and IMU. To track hand poses, the key to sensing is capturing high-quality information on the detailed motion and shape of the hand and fingers. The most obvious sensor to track motion is the IMU, which has been explored extensively and used by prior work to recognize discrete hand gestures by learning the fingerprints of the motion on the wrist resulting from performing different hand gestures \cite{GestureCustomization22Xu, Serendipity16Wen, tsunoda2024thumb}. However, the motion of the wrist does not contain enough information to infer the postures of all 20 finger joints. A more capable sensing modality is needed, one that can capture richer information on finger articulation. Recent work~\cite{Lee24_echoWrist} has shown that two pairs of microphones and speakers with specific configurations on a wristband can be used as micro-sonars to capture and infer accurate hand poses. Building on this insight, we hypothesize that \textit{repurposing the built-in microphone and speaker as a micro sonar can provide sufficient information to continuously track finger joint movements and hand poses on commercial smartwatches.}

\textcolor{blue}{To explore this idea, we implemented a system that leverages active acoustic sensing with a built-in speaker and microphone on COTS smartwatches to capture fine-grained finger movements. Our system emits encoded acoustic waves, and the microphone captures the reflected and diffracted signals, essentially functioning as an airborne sonar system. The emitted acoustic waves propagate through the air toward the hand; when they encounter the hand, a portion of the energy is reflected back toward the source, specifically the microphone, as an echo. The system then measures the time elapsed between the emission and the reception of the echo. To enhance the ranging resolution and continuously capture the complex spatial and temporal reflection strengths around the hand, we employed the C-FMCW technique~\cite{wang2018c}, as demonstrated in prior work \cite{Lee24_echoWrist, Yu24_ringAPose}. This technique estimates the round-trip time of the acoustic signal by identifying the correlation between the transmitted and received signals. The resulting spatiotemporal reflection patterns, encoding hand geometries, are subsequently used to infer hand poses. More detailed signal processing is described in Section \ref{acoustic-preprocessing}.}

\subsection{Data Collection System}
\label{watch_system}
We implemented a sensor data capture system on three different COTS smartwatches running on WearOS: Samsung Galaxy Watch 7, Xiaomi Watch 2 Pro, and Google Pixel Watch 3. These devices feature 425 mAh, 495 mAh, and 420 mAh Li-ion batteries, respectively. All three smartwatches feature a left-side speaker and a right-side microphone when worn on the right dorsal wrist and support simultaneous 48 kHz, 16-bit PCM audio playback and recording~\cite{Kim25OpenAcoustics}. While the speaker is consistently located at the center-bottom of the left side across all three models, the microphone placement varies slightly---centered on the right side in the Galaxy Watch, and shifted slightly toward the left and bottom on the Xiaomi and Pixel watches (see Figure~\ref{fig:watches}). The position will be opposite if the watch is worn on the left wrist. Different sensor positions will significantly impact the paths along which acoustic signals are reflected, a factor we explicitly explore in this paper. 

Our system emits FMCW signals in the inaudible range (18–21 kHz) and simultaneously captures reflected and diffracted signals from hand movements. To characterize the output levels across devices, we measured the playback sound levels using a sound level meter positioned 10 cm from the speaker. The Galaxy Watch 7 emitted 61.6 dBA, the Xiaomi Watch 2 Pro produced 66.9 dBA, and the Pixel Watch 3 emitted the highest amplitude at 80.8 dBA. To ensure consistently characterized signals across different hardware, our system records \textit{UNPROCESSED} signals using the Android MediaRecorder API~\footnote{\url{https://developer.android.com/media/platform/mediarecorder}}. Additionally, we recorded 6-axis IMU data at the maximum available sampling rate on each device (Galaxy Watch: 100 Hz; Xiaomi Watch: 50 Hz; Pixel Watch: 200 Hz) to explore its potential contribution to hand pose tracking.


\subsection{Acoustic Data Preprocessing}
\label{acoustic-preprocessing}

\subsubsection{C-FMCW Based Echo Profiles}
\textcolor{blue}{We employ airborne active acoustic sensing based on the C-FMCW approach~\cite{wang2018c}. In the conventional Linear FMCW method, the theoretical distance estimation, or range-finding, resolution ($R_{L-FMCW}$) is restricted by the sweep bandwidth ($B$) under the speed of sound in air ($C = 343 m/s$). Using our operational bandwidth, the resolution would be 57.2 mm ($= \frac{C}{2B} = \frac{343 m/s}{2 \times (21000 - 18000)} = 0.0572 m$) \cite{stove1992linear}. However, by leveraging the C-FMCW technique, the effective theoretical distance estimation resolution 
($R_{C-FMCW}$) is instead restricted by the system's sampling rate ($F_s$), allowing for a substantial increase in spatial precision: }

\textcolor{blue}{
\begin{equation}
    R_{C-FMCW} = \frac{C}{2F_s} = \frac{343 m/s}{2 \times (48000)} = 0.00357 m = 3.57 mm
\end{equation}
}

\begin{algorithm*}
\caption{\textcolor{blue}{Echo Profile Generation}}
\label{alg:echo_profile}
\begin{algorithmic}[1]

\Require 
    $Rx$: Raw received audio signal in the full recording $(N_{samples}, )$;
    $N_{samples}$: The length of $Rx$;
    $Tx$: Reference transmitted FMCW signal of one chirp, $(L, )$;
    $L$: Frame length, the samples per chirp, which is $(600)$;
    $F_s$: Sampling rate, which is $(48000)$;
    $[f_{min}, f_{max}]$: Bandpass frequency range, the frequency range of the transmitted signal, which is $(18000, 21000)$;
    $N_{order}$: Butterworth bandpass filter order, which is $(5)$
    $N_{frames}$: The number of total echo frames.

\Ensure 
    $P$: Original Echo Profiles $(L, N_{frames})$;
    $P_{diff}$: Differential Echo Profiles $(L, N_{frames} - 1)$

    \Statex
    \Statex \textbf{Step 1: Data Formatting and Filtering}
    \State $Rx \leftarrow \textsc{ButterworthBandpassFilter}(Rx, f_{min}, f_{max}, F_s, N_{order})$
    \Comment{Filter to match transmitted signal bandwidth}
    
    \Statex
    \Statex \textbf{Step 2: Start Position Detection}
    \State $Corr_{sync} \leftarrow Rx \star Tx$ 
    \Comment{$\star$ denotes Cross-Correlation}
    \State $p_{\text{start}} \leftarrow \arg\max(|Corr_{sync}|)$ 
    \Comment{Find index of strongest correlation}
    \State $p_{\text{start}} \leftarrow (p_{\text{start}} + L - \lfloor L/2 \rfloor) \mod L$ \Comment{Adjust to frame center}
    \State $Rx \leftarrow Rx[p_{start} :, :]$ 
    \Comment{Align to start position, $Rx$ shape: $(N_{samples} - p_{start}, )$}

    \Statex
    \Statex \textbf{Step 3: Original Echo Profile Computation}
    \State $y \leftarrow Rx \star Tx$ 
    \Comment{$\star$ denotes Cross-Correlation, $y$ shape: $(\text{Length}(Rx) + L - 1, )$}
    \State $N_{frames} \leftarrow \lfloor \text{Length}(y) / L \rfloor$
    \Comment{Calculate $N_{frames}$ with $L$}
    \State $P \leftarrow \textsc{Reshape}(y, [N_{frames}, L])$
    \Comment{Reshape into frames}
    \State $P \leftarrow (P)^T$ 
    \Comment{Transpose to (Distance, Time), $P$ shape: $(L, N_{frames})$}

    \Statex
    \Statex \textbf{Step 4: Differential Echo Profile Computation}
    \For{frame $f = 1$ to $N_{frames}$}
        \State $P_{diff}[f] \leftarrow |P[f]| - |P[f-1]|$ 
        \Comment{Calculate frame-to-frame magnitude difference}
    \EndFor
    \Comment{$P_{diff}$ shape: $(L, N_{frames}-1)$}

\end{algorithmic}
\end{algorithm*}

\textcolor{blue}{This marked improvement in resolution is critical for the fine-grained hand pose tracking required by our system. Prior work has shown that C-FMCW can support continuous hand pose tracking with a single speaker-microphone pair on a custom ring \cite{Yu24_ringAPose}. We apply this technique in our system and detail the signal processing pipeline in Algorithm~\ref{alg:echo_profile}. Once activated, our system continuously emits and records FMCW signals in the 18-21 kHz band.  The received signals are first passed through a 5th-order Butterworth bandpass filter with cutoff frequencies that match our emission band to eliminate out-of-band noise. Since the direct path between the speaker and microphone (without reflection) represents the shortest propagation path, and thus the earliest arrival, we use it as a reference for absolute distance. To locate this direct-path arrival, we calculate the cross-correlation between one transmitted FMCW sweep and the filtered received signals. The timestamp of the maximum correlation peak identifies the start of the received sweep, corresponding to the direct path, and serves as a reference for subsequent signal alignment. We then process the received signals in segments corresponding to each FMCW sweep, where each sweep consists of 600 samples ($L$). For every sweep, we calculate the cross-correlation between the transmitted and received signals. The resulting values encode reflection strength across different time lags, which we map to the distances using the speed of sound. This distance-mapped vector forms a slice of our echo profile, which we call an \textit{echo frame}. Temporally stacking these slices forms the complete \textit{original echo profile}. In the original echo profiles, each pixel along the x-axis represents the duration of one frequency sweep, which is 12.5 ms ($=  \frac{L}{F_{s}} = \frac{600}{48000} = 0.0125 s$). Along the y-axis, which consists of 600 pixels representing the sensing range, each pixel corresponds to a distance of approximately 3.57 mm \cite{wang2018c,Yu24_ringAPose}. The intensity of each pixel represents the strength of the correlation at that specific time/distance. }

\begin{figure*}[t]
\centering
   \includegraphics[width=\textwidth]{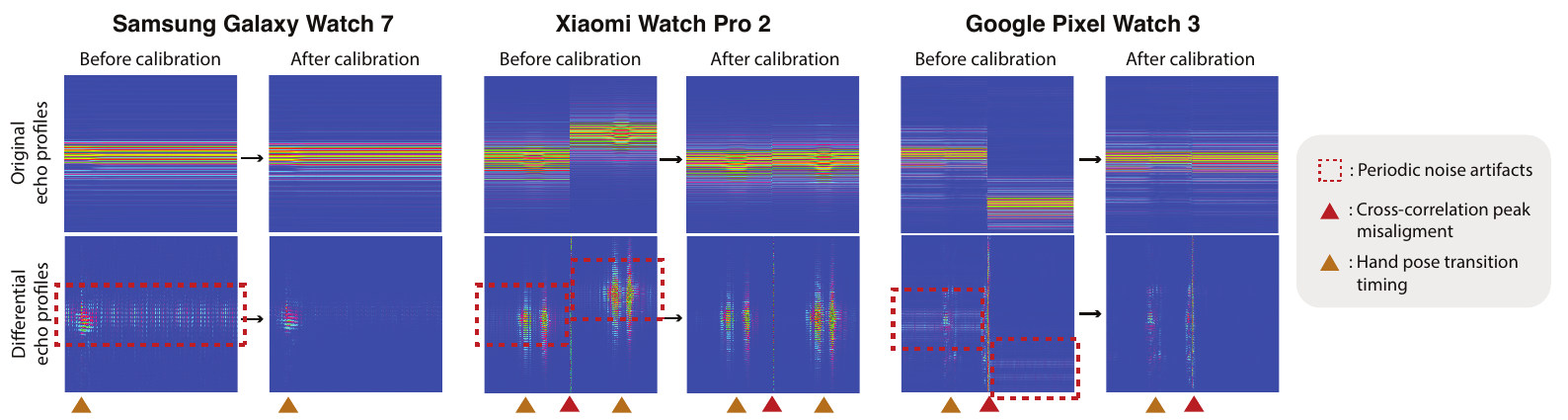}
   \hfil
\caption{Visual examples of echo profile calibration \textcolor{blue}{across three commercial smartwatches. Sliding-window cross-correlation peak correction removes peak misalignment (red triangles), while periodic drift calibration mitigates repeating noise artifacts (red dashed boxes). Hand pose transition timings are marked (yellow triangles). Together, these calibration steps produce cleaner and more temporally stable echo profiles during dynamic hand pose transitions.}}
\Description{Visual examples of echo profile calibration across three commercial smartwatches. Sliding-window cross-correlation peak correction removes peak misalignment (red triangles), while periodic drift calibration mitigates repeating noise artifacts (red dashed boxes). Hand pose transition timings are marked (yellow triangles). Together, these calibration steps produce cleaner and more temporally stable echo profiles during dynamic hand pose transitions.}
\label{fig:calibration}
\end{figure*}

\subsubsection{Differential Echo Profiles}
\textcolor{blue}{The original echo profiles preserve the hand shape information through the reflection strengths over time and distance, but they also include static reflections from the nearby surfaces and environments. To suppress these static components and highlight motion, we introduce \textit{differential echo profiles} by computing the frame-to-frame differences in the original echo profiles along the time axis, thereby capturing the temporal variations in acoustic reflections. Specifically, we subtract each preceding echo frame from the current one. This process helps attenuate static reflections while emphasizing moving elements, enhancing the system's ability to track finger movements.}

\textcolor{blue}{To focus on hand reflections and reduce environmental interference, we only retain the 60 closest pixels along the vertical axis of the echo profiles, corresponding to a sensing range of 21.42 cm. This range covers the typical wrist-to-middle-fingertip distance (average male: 19.05 cm, female: 18.29 cm \cite{tilley2001measure}). Along the horizontal axis, we use a 1.2-second window, which corresponds to 96 pixels.}
\textcolor{blue}{Altogether, combining the original and differential echo profiles, this results in two channels of $60 \times 96$ acoustic input to our models (see Figure \ref{fig:teaser}-B). The echo profiles within each window are normalized prior to being fed into the model to ensure consistency across different conditions.}

\textcolor{blue}{To enhance signal clarity and highlight salient patterns, we clip echo profiles to a threshold of $\pm 10^{10}$ for visualization. This threshold is determined empirically to best illustrate the finger motion patterns. All visualized echo profile figures in this paper use these clipped versions, offering a clearer view of the extracted features. However, to preserve maximum information for the deep-learning model, we use the raw unclipped echo profiles during training.}

\subsubsection{Echo Profile Calibration for Different Watches}
Ideally, the strongest cross-correlation peak, representing the fixed direct path between the smartwatch's microphone and speaker, should appear at a consistent position in each echo frame in the same recording. Reflections, inherently weaker, would then be analyzed relative to this point. However, our pilot study revealed a noticeable drift in the position of this dominant peak across consecutive echo frames in some recordings, especially for the Xiaomi and Pixel Watch (see Figure~\ref{fig:calibration}). We attribute this to device-specific timing jitter in the audio stack: the intermittent delay between chirp repetitions. \textcolor{blue}{This misalignment can shift echo profiles to incorrect ranges, disrupting the pattern, and leading the model to make entirely wrong predictions.} To resolve this problem, we adopted a simple yet practical approach, a sliding-window-based correction method. This method dynamically estimates the peak location within overlapping windows and realigns segments exhibiting substantial shifts. 

A further challenge was a periodic drift of echo frames within the original echo profiles. This drift appeared as periodic artifacts in the differential echo profiles. Several factors may contribute to this issue, including subtle periodic variations in the effective sampling rate and modulated aliasing effects. To mitigate these periodic artifacts, we implemented a median filter on the original echo profiles. \textcolor{blue}{The second row of} Figure~\ref{fig:calibration} shows that this filtering approach effectively eliminates the subtle drifts while preserving temporal consistency in the original data, consequently removing the artifacts from the differential echo profiles and ensuring cleaner signal representation.

\textcolor{blue}{We found that the Xiaomi and Pixel Watch exhibit more misalignment issues, while the Galaxy Watch shows more drifting issues. These device-specific inconsistencies in echo profiles can significantly degrade performance, especially when training on data pooled across different brands. Our calibration process can effectively address these issues with a universal, hardware-agnostic algorithm that corrects these misalignment and drifting issues. Importantly, all echo profiles from all devices are processed using the identical calibration process, with no per-watch or per-model specialization.}

\subsection{IMU Motion Data Preprocessing}
\label{imu-preprocessing}
To enhance feature extraction from IMU data while reducing noise, we adopted a data preprocessing method from prior work~\cite{GestureCustomization22Xu}. The raw 6-axis IMU data, comprising 3 axes of accelerometer data and 3 axes of gyroscope data, yields 6 initial data channels. Each channel undergoes processing through a 2nd-order Butterworth bandpass filter with three cutoff frequency ranges: 0.22-8 Hz, 8-32 Hz, and above 32 Hz. This filtering approach separates different frequency components of the signal. By preserving the original raw data and adding these three filtered versions, each original channel expands into 4 channels. Consistent with the acoustic data processing, we employ a window of 1.2 seconds for the IMU data, and the data within each window is normalized prior to being fed into the model.

\begin{figure*}[t]
\centering
   \includegraphics[width=\textwidth]{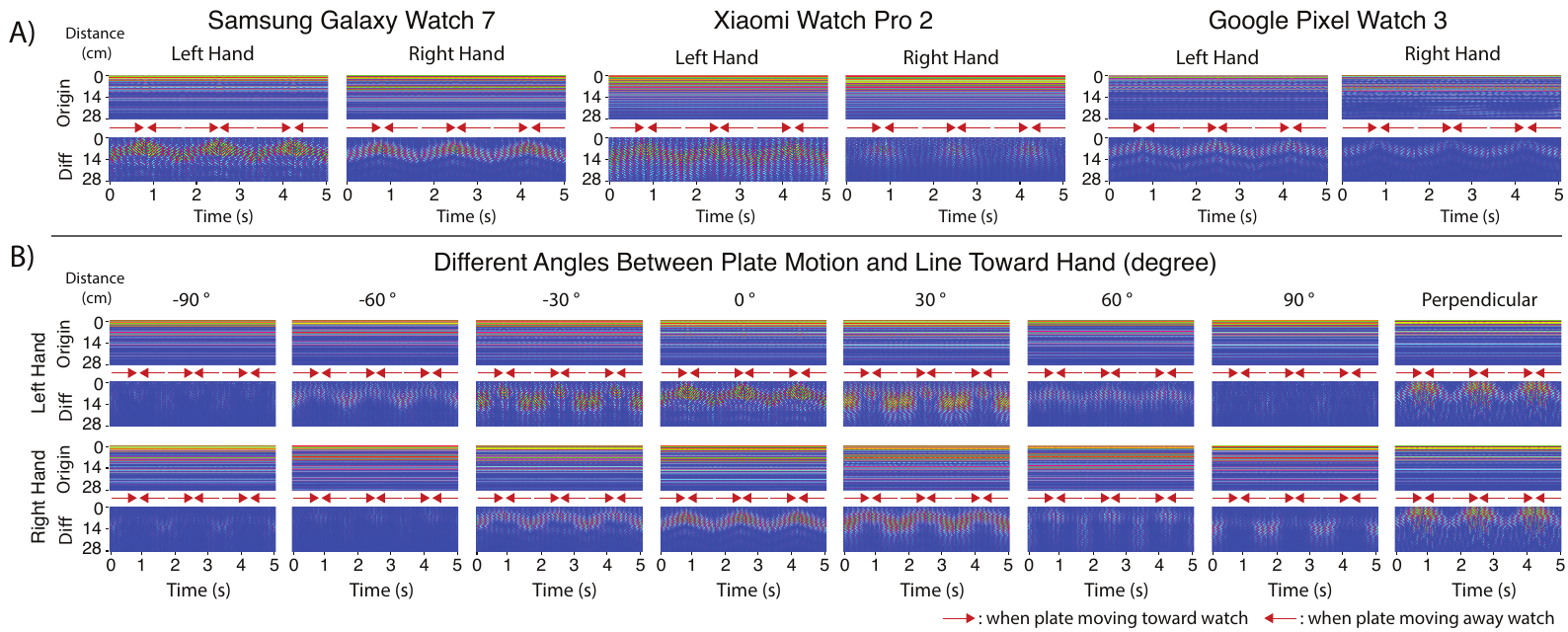}
   \hfil
\caption{Captured original and differential echo profiles across different COTS smartwatches (Galaxy, Xiaomi, and Pixel) (A) and varying angles between the plate's motion and line toward the hand (B) during repetitive back-and-forth movements toward and away from the smartwatch.}
\Description{Figure 4 presents captured echo profiles from different COTS smartwatches—Samsung Galaxy Watch 7, Xiaomi Watch Pro 2, and Google Pixel Watch 3—under two experimental conditions. In (A), echo profiles (both original and differential) are shown for left- and right-hand wearing configurations, illustrating consistent motion patterns across hardware and orientations. In (B), the figure shows echo profiles at varying angles (from –90° to +90°, and perpendicular) between the direction of a moving plate and the line toward the hand. These results highlight how motion direction and hand orientation influence the structure of acoustic reflections, demonstrating the system’s sensitivity to 3D movement cues.}
\label{fig:moving_plate_echoprofile}
\end{figure*}

\subsection{Preliminary Viability Testing on COTS Watches}

To demonstrate the viability of our system for 3D hand pose tracking on COTS smartwatches, we visually examined processed data to assess how hardware configurations and watch-wearing hands affect echo profiles, whether a single built-in speaker–microphone pair can reliably capture 3D around-device movement, and how echo profiles can support reliable 3D hand pose tracking.

\subsubsection{Impact of hardware and watch-wearing hand}
We first investigated how differences in smartwatch hardware and the wearing hand affect captured echo profiles. To simulate consistent around-device movement, we used a stepper motor-driven linear stage to move a flat plate back-and-forth at 6 cm/s over a prop hand wearing a smartwatch, within a 10–15 cm range—an area corresponding to typical finger movement (see Figure~\ref{fig:watches}-B). Visual inspection of the resulting differential echo profiles (Figure~\ref{fig:moving_plate_echoprofile}-A) shows that while all watches reliably capture distinct motion patterns, the signal characteristics vary across devices and hand orientations. For instance, the Galaxy Watch exhibits clear zigzag patterns, the Xiaomi Watch produces broader profiles, and the Pixel Watch shows narrower responses. Differences in signal intensity were also observed between left- and right-hand configurations (which flip the watch, and thus the orientations of the microphone and speaker). This highlights the importance of accounting for real-world variability in both hardware and watch-wearing hands.

\subsubsection{3D Around-Device Movement Tracking Viability}
Next, we investigated the capability of active acoustic sensing on a COTS smartwatch to support 3D around-device movement tracking. While SonarSelect~\cite{kim2025cross} demonstrated robust sensing of one-dimensional motion using sonar on a COTS smartwatch, other recent studies~\cite{SonarID22Kim, Zhang24LipWatch} have shown that more sophisticated around-device movement (e.g., lip motions for silent speech recognition) can be detected using only a single built-in speaker and microphone. To evaluate this capability in the context of 3D hand pose tracking, we visually examined how changes in movement direction affect the captured echo profiles. Specifically, we varied the angle between the motion path of a moving plate and the long axis of the hand/wrist from –90° to 90° in 30° increments and additionally tested a perpendicular movement angle involving motions towards and away from the palm (see Figure~\ref{fig:watches}-B). Figure~\ref{fig:moving_plate_echoprofile}-B illustrates the results. It shows distinct patterns corresponding to different around-movement angles. For example, while a clear straight zigzag pattern appears when the movement is directed toward the hand (0°), as the angle changes, the echo profiles vary, showing discontinuities at ±30° on the left-hand, curved signals at ±30° on the right-hand, signal shifts when the movement is perpendicular to the watch screen, and weaker signals when the angle exceeds 60° for both hands. These observations highlight that the captured echo profiles contain rich 3D around-device movement information, which can potentially be leveraged to enable continuous 3D hand pose tracking.

\begin{figure*}[t!]
\centering
   \includegraphics[width=\textwidth]{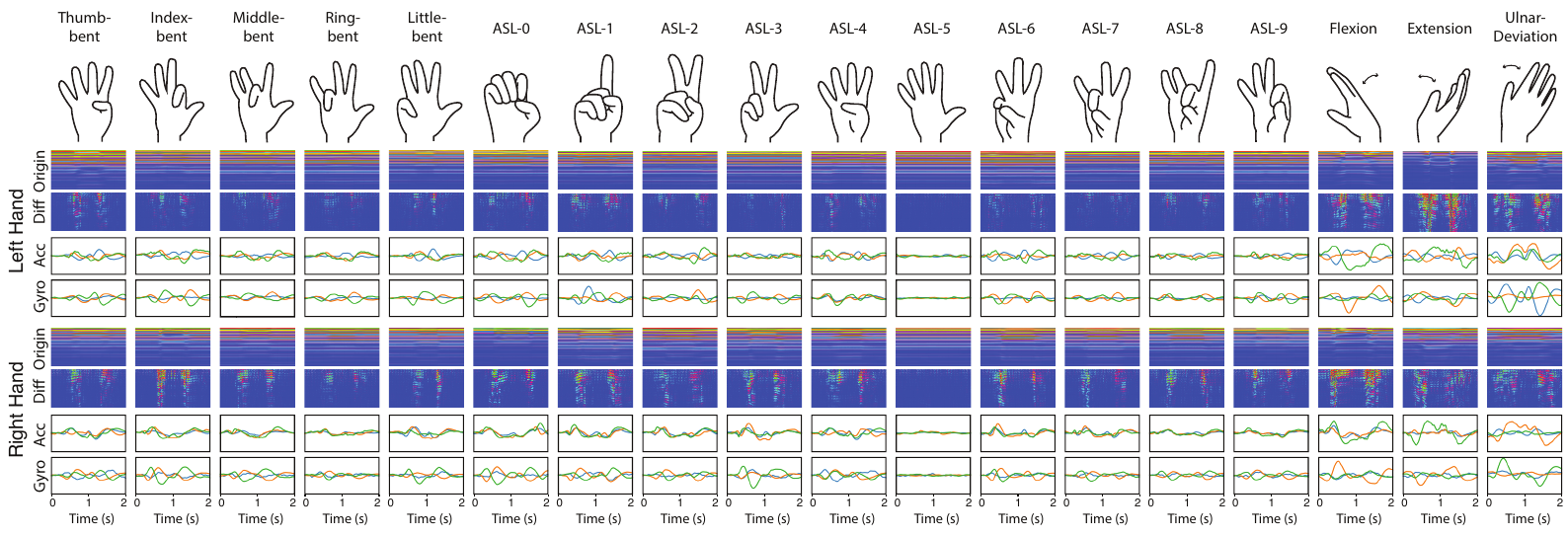}
   \hfil
\caption{Processed original and differential echo profiles (see Section~\ref{acoustic-preprocessing}) and bandpass-filtered IMU sensor data in the 32-100 Hz range (see Section~\ref{imu-preprocessing}) captured from a single user using a COTS smartwatch (Galaxy Watch 7) during different hand postures.}
\Description{
Figure 5 shows processed original and differential echo profiles, along with bandpass-filtered IMU sensor data (32–100 Hz), for various hand postures captured from a single user wearing a COTS smartwatch (Galaxy Watch 7). Data are shown separately for left- and right-hand wearing conditions across a range of hand poses. The figure illustrates how different hand configurations produce distinct acoustic and motion signal patterns, highlighting the system’s ability to capture fine-grained variations in hand articulation.
}
\label{fig:data_vis}
\end{figure*}

\subsubsection{3D Hand Pose Tracking Viability}
Given these insights, we further explored the viability of continuous hand pose tracking across different hand configurations. Figure~\ref{fig:data_vis} presents samples of processed acoustic and motion data recorded by our system for various hand poses performed by a single user. Each pose was executed by transitioning from an open-hand position to the target pose and then returning to the open-hand position. A visual examination of the sensor data indicates that different hand poses generate sufficiently distinct patterns, supporting the potential for continuous hand pose tracking. Most poses exhibit noticeable variations, except for ASL-5, where the hand remains open with minimal movement. Additionally, wrist rotation poses produce clearer patterns in the differential echo profiles. Collectively, these observations confirm the viability of 3D hand pose tracking on COTS smartwatches using only their built-in sensors.

\subsection{Deep-Learning Pipelines}
\label{deep-learning pipeline}

\begin{figure}[t]
\centering
   \includegraphics[width=8.5cm]{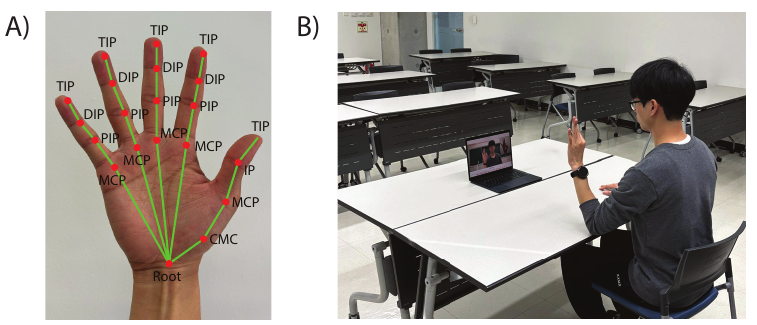}
   \hfil
\caption{\textcolor{blue}{(A) MediapPipe-based 3D hand joint annotation used as ground truth, showing 20 labeled landmarks from the root to each fingertip. (B) Study setup in which the laptop’s front camera faces the participant’s palm to capture ground truth without optical occlusion during hand pose demonstrations.}}
\Description{Figure 6-A shows MediapPipe-based 3D hand joint annotation used as ground truth, showing 20 labeled landmarks from the root to each fingertip. Figure 6-B illustrates Study setup in which the laptop’s front camera faces the participant’s palm to capture ground truth without optical occlusion during hand pose demonstrations.}
\label{fig:study_setup}
\end{figure}

\subsubsection{Ground Truth Acquisition \& Normalization}
\textcolor{blue}{Since our sensing technique is sensitive to the physical geometry of the hand, relying on marker-based systems potentially distorts the signals. To establish reliable ground truth for continuous 3D hand pose information without altering the hand's acoustic reflective properties, we employ MediaPipe Hands~\cite{zhang2020mediapipe}.} \textcolor{red}{Although monocular RGB cameras can suffer from reduced accuracy when finger joints are occluded, we minimized such occlusions by positioning the camera to face the user’s palm directly (see Figure~\ref{fig:study_setup}).} \textcolor{blue}{This RGB camera-based approach has been widely adopted by prior work in wearable hand tracking for generating ground truth~\cite{Lee24_echoWrist, Yu24_ringAPose, waghmare2023z, Kim22EtherPose, Devrio22DiscoBand}. Hand poses are captured at 30 frames per second (FPS) using a built-in webcam on a MacBook Air.} Processing webcam-captured video footage with MediaPipe Hands produces 3D hand representations consisting of 21 hand joint coordinates in 3D space. We designate the wrist coordinate as the reference point (origin) of our coordinate system and calculate the relative positions of the remaining 20 landmarks by subtracting the wrist coordinate from each point. This approach provides a consistent frame of reference centered on the wrist position. We normalize the 3D hand pose ground truth to address session-to-session and user-to-user differences in hand position, orientation, and size. For each generated frame, we first define the palm plane using vectors originating from the wrist and extending to the metacarpophalangeal (MCP) joints of the index and little fingers. We then calculate a rotation matrix to align this palm plane with a reference palm plane derived from a baseline posture. This rotation correction compensates for the inevitable variations in hand orientation that occur between different recording sessions and when users remount the device. Additionally, for each user, we normalize the hand size by scaling each frame so that the distance between the wrist and the little finger MCP joint matches a measured physical length. Note that the wrist rotation is not normalized to preserve its natural variation. This normalization method ensures that generated hand pose data remains comparable across different recording sessions and users.

\subsubsection{Deep-learning Model Framework}    \label{sec:acoustic-model-framework}
We construct a regression model that learns the mapping from acoustic data to 3D hand poses. Each ground truth video frame serves as an individual sample at \textcolor{blue}{30} FPS, with echo profiles mapped to ground truth using timestamp alignment. \textcolor{blue}{To support continuous hand pose tracking, all the frames are used, including the transition segments.} We employ the FastViT-T12~\cite{vasu2023fastvit}, a state-of-the-art hybrid model that combines the efficiency of convolutional neural networks (CNNs) with the powerful feature mixing capabilities of vision transformers. \textcolor{blue}{In this process, we also explored ResNet-18~\cite{he2016deep} and CNN-LSTM that have been explored in closely related prior works~\cite{Lee24_echoWrist, Yu24_ringAPose, lee2025grab}, and we chose the best-performing and most lightweight model.} The model takes a \textcolor{blue}{two-channel tensor as input, comprising original and differential echo profiles}. The architecture processes this input through a convolutional stem for initial $4x$ downsampling, followed by four hierarchical stages that progressively refine features using RepMixerBlocks. Between stages, a PatchEmbed layer further downsamples the feature maps. A final regression head uses global average pooling and a linear layer to map the learned features to the 60 output parameters representing the 3D coordinates of 20 hand landmarks (see Figure~\ref{fig:teaser}). \textcolor{blue}{These landmarks can be used directly to reconstruct the wireframe of the hand as shown in Figure~\ref{fig:study_setup}. We subsequently generate all the visualizations using MANO~\cite{Romero17MANO}.} The model is trained end-to-end using a composite loss function that promotes both positional accuracy and temporal smoothness. This loss is a weighted sum of the standard Mean Squared Error (MSE) between the predicted and ground truth coordinates, and a velocity loss. The velocity component, weighted by a factor of 0.1, penalizes abrupt changes between consecutive frames, encouraging the model to produce more physically plausible motion. \textcolor{blue}{Overall, the model has 6.62M parameters.}

To explore whether IMU signals could complement acoustic sensing, we extended our pipeline with a lightweight IMU encoder. Inspired by prior work on IMU-based full-body pose tracking~\cite{mollyn2023imuposer}, filtered motion signals (see Section~\ref{imu-preprocessing}) were first mapped to a 512-dimensional embedding via a linear layer, then processed using a bidirectional LSTM. The resulting features were concatenated with the acoustic features and passed through a simple fusion decoder to predict the 3D hand landmarks.

\subsubsection{Training Pipeline}   
\label{sec:training-pipeline}
We use three protocols to train and evaluate our models:

\begin{itemize}
    \item Cross-user models: We train cross-user foundation models using data from all participants except one, whose data was reserved exclusively for the test set. This protocol simulates the deployment scenario where a system is pre-trained on large-scale datasets without access to the end user's data.
    \item Cross-session models: User-specific models created by fine-tuning cross-user foundation models with target user data from all wearing sessions except the final session, which is reserved for testing. This protocol mimics realistic usage where users provide personalization data before regular system deployment.
    \item Within-session models: User-specific models that fine-tune cross-user foundation models using data from sessions up to the antepenultimate session plus a 50\% subset from the final two sessions. The remaining 50\% of data from the final two sessions serves as the test set. This protocol evaluates system performance when users collect calibration data after watch remounting, simulating scenarios where brief personalization occurs with every device repositioning.
\end{itemize}

We employed a consistent two-phase fine-tuning training strategy for the acoustic model to maximize the utility of our collected dataset. In the first phase, we develop a cross-user foundation model that captures generalizable hand pose tracking patterns across participants. This cross-user base model is trained using the Adam optimizer with an initial learning rate of 0.001, incorporating a learning rate scheduler to dynamically adjust optimization during training. Training continues for 100 epochs with early stopping if the validation metric does not improve for 9 epochs, utilizing a relatively large batch size of 256 to promote stable convergence. The second phase is user-specific cross-session fine-tuning. Starting with the pre-trained cross-user base model, we adapt the network to individual hand characteristics using user-specific data. This fine-tuning process employs the Adam optimizer with a reduced initial learning rate of 0.0003, allowing for more subtle parameter adjustments. We train for 100 epochs with a smaller batch size of 64, enabling the model to better capture user-specific nuances. This two-phase approach leverages our complete dataset efficiently, establishing broad hand-tracking capabilities in the base model while accommodating individual variations through targeted fine-tuning.

\subsubsection{Data Augmentation}
To enhance model robustness towards interference noise and mitigate overfitting, we implement data augmentation during training. While the original data is preserved, we introduce controlled randomness into copied data samples. For acoustic data, we apply three augmentation methods targeting different aspects of signal variability. First, we introduce random vertical shifts to the echo profiles, simulating variations in watch-wearing positions across different users and sessions. Specifically, we apply random shifts of $\pm 5$ pixels, \textcolor{blue}{corresponding to around 1.8 cm ranges to account for these natural positioning inconsistencies across users and watch-wearing sessions and enhance generalizability.} 
Second, we incorporate amplitude variations to increase resilience to signal strength fluctuations. With 80\% probability, we multiply echo profile values by a random factor between 0.95 and 1.05 ($\pm 0.5\%$). Third, to further challenge the model's ability to extract robust features, we applied time and frequency/distance masking. With 80\% probability, we mask random sections: 5 to 15 pixels on the x-axis (time) and 5 to 10 pixels on the y-axis (distance). This introduces controlled randomness into the training process, challenging the model to maintain performance despite minor signal variations rather than memorizing exact patterns.
\section{Study 1: Continuous Hand Pose Tracking} \label{sec:study}
We conducted a study to evaluate the continuous 3D hand pose tracking performance of WatchHand across three different COTS smartwatches and different watch-wearing hands. This study was approved by the local Institutional Review Board (IRB).

\subsection{Participants}
We recruited 24 participants (12 males, a mean age of 22.5 (SD 2.3)) from the local university. 23 participants were right-handed. While most participants reported wearing a smartwatch on their non-dominant hand, two right-handed participants preferred wearing it on their dominant hand for reasons such as easier access to notifications or increased comfort. Participants’ hand sizes averaged 17.95 cm (SD 0.95) in length (from the base of the hand to the tip of the middle finger) and 7.88 cm (SD 0.53) in palm width. The study took approximately two hours to complete, and participants were compensated with 30 USD in local currency. 

\subsection{Study Design}
We followed an established methodology previously tested on customized hardware~\cite{Lee24_echoWrist} to ensure the comparability of our results. Our study evaluated a total of 18 gestures (see Figure~\ref{fig:data_vis}), categorized into three groups: (1) simple gestures (five single-finger bending movements), (2) complex gestures (ten complex finger gestures representing American Sign Language (ASL) digits 0–9), and (3) wrist orientations (three distinct wrist movements: flexion, extension, and ulnar deviation). \textcolor{blue}{Although the thumb-bent and ASL-4 poses represent similar poses and appear in both the simple and complex gesture sets, we retained them to enable direct comparison with prior work~\cite{Lee24_echoWrist}. Since our system performs continuous per-frame hand pose tracking rather than discrete gesture classification, this duplication has minimal impact on the validity of our results.} Each session consisted of four repetitions per gesture, with each gesture lasting two seconds. \textcolor{blue}{We randomized the gesture order within each group to minimize potential order effects, but kept the group order (simple → complex → wrist rotation) fixed to reduce participant confusion. This design choice follows prior work~\cite{Lee24_echoWrist}, enabling direct comparison with their results.} While prior work conducted 20 sessions per participant using only a single hand, our study introduced two independent variables: hardware (Samsung Galaxy Watch 7, Xiaomi Watch 2 Pro, Google Pixel Watch 3) and hand orientation (left, right). The hardware variable was between groups, while the hand orientation variable was within groups. We evenly distributed the 24 participants across the three smartwatch groups, with eight different participants performing all study tasks on each smartwatch. Each participant completed 20 sessions, with 10 sessions per hand, to ensure comprehensive data collection across different conditions. The order in which participants completed trials with the watch on their right and left hands was fully counter-balanced within each watch group.

\subsection{Study Procedure}
\label{procedure}
The study was conducted in an empty room with participants seated at a desk. To prevent physical fatigue, participants rested their elbows on the desk. Participants first signed an informed consent form, had their hand size measured, and read the experimental instructions. Before starting the experiment, they practiced all 18 different hand poses for five minutes. If some poses could not be performed due to ergonomic constraints, participants were instructed to approximate the pose. For example, 13 participants were unable to bend their little finger alone, so they bent the ring finger together (8) or rotated it toward the palm (5). Next, participants donned the smartwatch on the instructed wrist, and the study began. For each session, the target pose appeared on a screen in front of the participant for 2 seconds (\textcolor{blue}{see Figure~\ref{fig:study_setup}-B}), and they were instructed to perform the pose and then return to the default pose (ASL-5). The next target pose would then appear. Each session lasted approximately 3 minutes. Participants were instructed to remove and re-wear the smartwatch between sessions and were able to rest between sessions. This re-wearing process simulates real-world use, as users often remove and re-wear smartwatches. After completing 10 sessions, they switched the smartwatch to the opposite wrist and conducted the remaining 10 sessions. At the end of the study, participants completed a demographic questionnaire.
\section{Results}

We recorded a total of 1152 minutes of gesture data from 24 participants, consisting of 192 minutes for each of the six combinations of two independent factors: three smartwatch models and two wearing hands. During the study, while some participants at times performed incorrect gestures due to physical constraints or loss of focus, we included all recorded data, as our ground truth capture system reliably tracks the actual gestures performed, and our interest lies in continuous hand pose tracking. These natural variations enrich the dataset’s diversity and contribute to the development of models that are more robust to continuous and unconstrained hand pose tracking scenarios.

To evaluate the performance of WatchHand for continuous hand pose tracking, we first segmented the collected gesture data into two distinct subsets, following prior work~\cite{Lee24_echoWrist, Kim22EtherPose}: (1) fine-grained finger movements, consisting of five simple gestures and ten complex gestures, and (2) gross wrist rotations, including three wrist rotation gestures. Then, we explored the performance result with respect to different smartwatch models, variations in watch orientation, and evaluation conditions (e.g., cross-session, within-session, and cross-user models). We processed all the acoustic and motion data following the procedures outlined in Sections~\ref{acoustic-preprocessing} and~\ref{imu-preprocessing}, then trained the models in six conditions (three watches by two watch-wearing hands). \textcolor{blue}{In this result, for the acoustic model, the input consists of two channels of echo profiles (original and differential), each of size 96×60, as illustrated in Figure~\ref{fig:teaser}-C. The multimodal model additionally includes 24 IMU channels (six axes × four bands: raw plus three filtered bands). Importantly, the model is trained and evaluated on all frames in each recording, including both the instructed poses and the transitions into and out of those poses. We use a 1.2-second sliding window with a stride of 1/30 second for each step, so that we obtain one input window per video frame (30 FPS). This design explicitly targets continuous hand pose tracking rather than static pose classification.}

As detailed in Section~\ref{sec:training-pipeline}, we implemented and evaluated cross-session, within-session, and cross-user models, following a two-phase training pipeline. In this analysis, as all the results were normally distributed and met the upheld-sphericity assumptions, we conducted two-way ANOVAs on both metrics with different watch hardware and watch-wearing hand as independent variables. 

\begin{figure*}[t]
\centering
   \includegraphics[width=\textwidth]{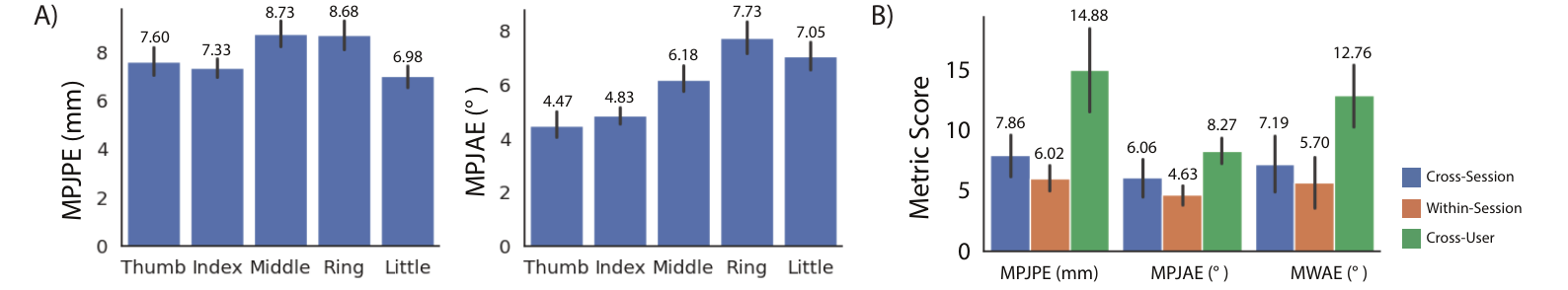}
   \hfil
\caption{Hand pose tracking performance in Study 1: (A) average 3D hand reconstruction errors in cross-session evaluation for each finger in both MPJPE and MPJAE metrics, and (B) average hand pose tracking performance across all evaluation models and metrics.}
\Description{Figure 7  provides a detailed performance analysis of WatchHand’s 3D hand pose tracking system. (A) breaks down cross-session tracking errors by finger, showing that the middle and ring fingers exhibit the highest MPJPE and MPJAE due to their anatomical proximity and occlusion, while the little and index fingers perform best. (B) compares overall tracking performance across within-session, cross-session, and cross-user evaluations, revealing a clear degradation trend as generalization increases. Despite this, WatchHand maintains low MPJPE (7.86 mm), MPJAE (6.06°), and MWAE (7.19°) in cross-session scenarios, confirming its robustness for real-world deployment with session and user variability.}
\label{fig:study1_result}
\end{figure*}

\subsection{Cross-Session Model}
\label{sec:cross-session}

\subsubsection{Continuous Fine-Grained Finger Movements}

We evaluated system performance using two standard metrics: mean per-joint position error (MPJPE) and mean per-joint angle error (MPJAE). Across all conditions, the acoustic model achieved an average MPJPE of 7.8\textcolor{blue}{7} mm (SD 1.\textcolor{blue}{6}7) and an MPJAE of 6.\textcolor{blue}{06}$^\circ$ (SD 1.\textcolor{blue}{52}). For both metrics, no significant main effects or interactions were found (F-values \textcolor{blue}{0.005}-0.\textcolor{blue}{42}, p-values  0.\textcolor{blue}{66}-0.\textcolor{blue}{95}). These results confirm that WatchHand can reliably track continuous 3D hand poses for fine-grained finger movements across varying hardware and watch-wearing hands, using only acoustic data. By breaking down MPJPE and MPJAE metrics, Figure~\ref{fig:study1_result}-A illustrates the average 3D hand reconstruction errors for each finger. While MPJPE errors generally correlate with finger length, MPJAE more clearly highlights the middle, ring, and little fingers as primary sources of angular error, corresponding to regions where unintended or co-activated movements most commonly occur (e.g., involving \textcolor{blue}{ring finger motion when instructed to bend the little finger solely, a physical constraint that varies across participants}).

\subsubsection{Continuous Wrist Rotation}
Next, we evaluated our system's wrist-rotation tracking performance using the mean wrist angle error (MWAE) metric. The acoustic model demonstrated high accuracy in wrist-rotation tracking, achieving an average MWAE of \textcolor{blue}{7.19}$^\circ$ (SD 2.\textcolor{blue}{26}). Similar to the results for fine-grained finger movement tracking, no significant main effects or interactions were observed (F-values 0.\textcolor{blue}{12}-\textcolor{blue}{2.37}, p-values 0.\textcolor{blue}{11}-0.\textcolor{blue}{73}). 

\subsubsection{Multimodal Model Result}

We then evaluated the multimodal model combining acoustic and IMU data, which achieved an average MPJPE of 7.9\textcolor{blue}{6} mm (SD 1.\textcolor{blue}{6}9), MPJAE of 6.\textcolor{blue}{03}$^\circ$ (SD 1.\textcolor{blue}{49}), and MWAE of \textcolor{blue}{7.48}$^\circ$ (SD 2.\textcolor{blue}{23}). Similar to the acoustic model, performance showed no significant effects of hardware or watch-wearing hand (F-values 0.\textcolor{blue}{02}-1.\textcolor{blue}{83}, p-values 0.\textcolor{blue}{17}-0.9\textcolor{blue}{6}). This indicates performance is stable and similarly robust across all three watches and on both left and right wrists. A one-way ANOVA comparing acoustic and multimodal models revealed a non-significant effect with IMU addition in all metrics (F-values 0.0\textcolor{blue}{1}-0.\textcolor{blue}{41}, p-values 0.\textcolor{blue}{52}-0.92), indicating inclusion of IMU data did not impact model performance. \textcolor{blue}{This may be attributed to redundancy between the acoustic and motion signals. IMU data primarily captures wrist motion and its dynamics, and thus lacks the granularity needed to represent detailed finger articulation. In contrast, active acoustic sensing can detect fine-grained finger movements while simultaneously encoding wrist motion through reflected signals~\cite{Lee24_echoWrist}. As a result, the acoustic channel alone already provides both local (finger) and global (wrist) information, limiting the additional benefit that IMU data can offer, rather than indicating overfitting of the acoustic model.} Based on this observation, we decided to omit further analysis of the IMU data in subsequent evaluations to avoid redundant analysis.

\subsection{Within-Session Model}
To enable full comparison with prior work, we further evaluated WatchHand using a within-session model. It achieved an average MPJPE of \textcolor{blue}{6.02} mm (SD \textcolor{blue}{1.04}), MPJAE of 4.\textcolor{blue}{63}$^\circ$ (SD 0.\textcolor{blue}{80}), and MWAE of 5.\textcolor{blue}{70}$^\circ$ (SD 2.\textcolor{blue}{07}). Similar to the cross-session model, performance showed no significant effects of hardware or watch-wearing hand (F-values 0.\textcolor{blue}{004}-\textcolor{blue}{1.51}, p-values 0.\textcolor{blue}{27}-0.\textcolor{blue}{95}). The hand pose tracking performances are notably better than those from the cross-session model (see Figure~\ref{fig:study1_result}-B). These results align with prior works showing that within-session testing benefits from stable device placement and minimized variability~\cite{Hu20FingerTrak, Devrio22DiscoBand, Kyu24EITPose}. While this protocol represents a somewhat unrealistic best-case scenario with highly stable device placement, it does provide a useful benchmark against prior systems that have followed similar procedures and serves to underscore the effectiveness of WatchHand’s acoustic pipeline.

\subsection{Cross-User Model}
We next further validate WatchHand's performance using a cross-user model with a leave-one-participant-out scheme, where each participant's data was excluded from training and used solely for testing. The acoustic model shows an average MPJPE of 1\textcolor{blue}{4.88} mm (SD 3.\textcolor{blue}{37}), MPJAE of 8.\textcolor{blue}{27}$^\circ$ (SD 1.\textcolor{blue}{04}), and MWAE of 1\textcolor{blue}{2.76}$^\circ$ (SD 2.\textcolor{blue}{56}). Similarly, no significant main or interaction effects were observed for MPJPE \textcolor{blue}{and MWAE} (F-values 0.\textcolor{blue}{09}-1.7\textcolor{blue}{8}, p-values 0.1\textcolor{blue}{8}-0.7\textcolor{blue}{7}). However, hardware effects emerged for MPJAE (F (2, 42) = \textcolor{blue}{4.76}, p = \textcolor{blue}{0.01}, $\hat{\eta}^2_G$=0.\textcolor{blue}{18}). Post hoc t-tests revealed that the Galaxy watch shows weaker prediction performance in MPJAE, with a mean of \textcolor{blue}{8.84}$^\circ$ (SD 0.\textcolor{blue}{80}), compared to the Xiaomi watch (\textcolor{blue}{7.77}$^\circ$ (SD 1.1\textcolor{blue}{0}), p = 0.00\textcolor{blue}{3}). While we observed significant hardware effects in cross-user evaluation, these hardware differences were not evident in cross-session or within-session evaluations. This suggests that hardware-induced variability primarily emerges when models are generalized across users, where differences in acoustic signal amplitude and articulation patterns interact with device-specific and user-specific characteristics. We note that performance markedly decreased in the cross-user model compared to the cross-session or within-session model (see Figure~\ref{fig:study1_result}-B). \textcolor{blue}{This pattern is consistent with prior wrist-worn continuous hand pose tracking systems~\cite{Lee24_echoWrist, 12digitsKim, Devrio22DiscoBand, Kyu24EITPose, Hu20FingerTrak}, and} likely due to the high inter-user variability in hand shape, finger articulation, and motion styles \textcolor{blue}{rather than model overfitting}. These results suggest the potential value of incorporating lightweight personalization techniques, such as few-shot fine-tuning, to bridge the performance gap between generalized and personalized models. 
\section{Study 2: Adaptation to Diverse Body Postures}   
\label{sec:follow-up-study}
\begin{figure*}[t]
\centering
   \includegraphics[width=\textwidth]{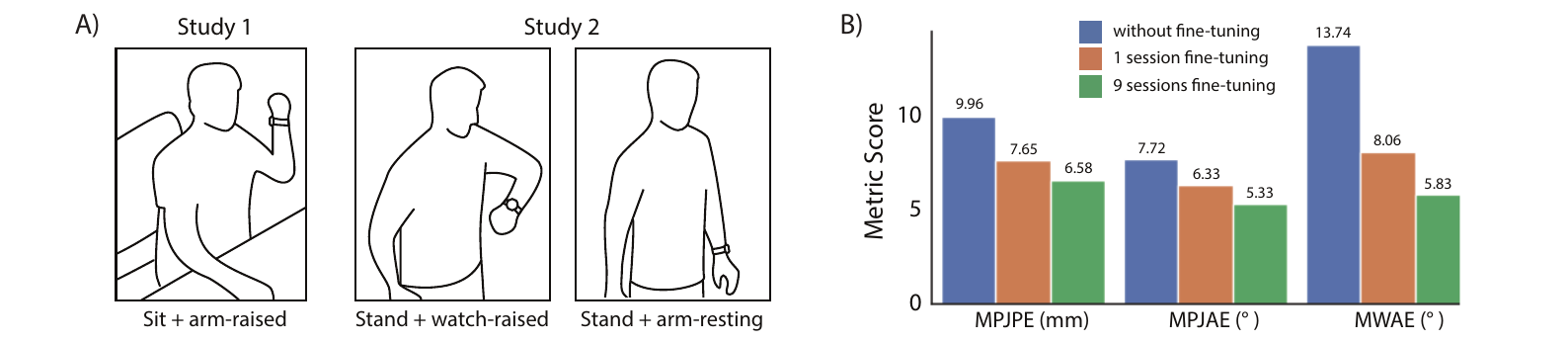}
   \hfil
\caption{(A) Three key postures covered in this work: sitting and arm-raised pose, and watch-raised pose, where the arm is naturally positioned horizontally in front of the body to view the watch screen, and arm-resting, with the arm lowered in a relaxed position adjacent to the thigh. (B) Average hand pose tracking performances from standing postures (including both watch-raised and arm-resting poses) for each fine-tuning condition---without (trained with data captured from the sitting poses), 1-session, and 9-session, reported across all metrics.}
\Description{Figure 8-A illustrates the three key postures evaluated in this work: (A) sitting with an arm-raised pose from the first study, (B) watch-raised pose where the arm is held horizontally in front of the body to view the watch, and (C) arm-resting pose with the arm relaxed beside the thigh. Postures (B) and (C) were assessed in the second study to examine the system’s robustness under natural variations in arm position. Figure 8-B shows the average hand pose tracking performance from standing postures (including both watch-raised and arm-resting poses) across three fine-tuning conditions: without fine-tuning, with 1-session fine-tuning, and with 9-session fine-tuning. Results across all metrics (MPJPE, MPJAE, MWAE) demonstrate that performance improves as more fine-tuning data is used, highlighting the benefit of adaptation for enhancing model accuracy in varied postural contexts.}
\label{fig:study2_result}
\end{figure*}

Given the positive hand pose tracking accuracy observed in sitting posture during Study 1, we further conducted a follow-up study to evaluate performance across more diverse postures. Specifically, we examined two common standing postures, including a \textit{watch-raised} pose and an \textit{arm-resting} pose (Figure~\ref{fig:study2_result}-A). This study was approved by the local IRB. 

\subsection{Participants}
We re-invited six participants (three males, a mean age of 22.17 (SD 2.86), all right-handed) from the first study, sampled opportunistically according to their availability. Each participant wore the same smartwatch model they had previously used (3 Xiaomi Watch, 2 Google Pixel Watch, and 1 Samsung Galaxy Watch) on their non-dominant hand. This study lasted approximately two hours, and each participant was compensated with 30 USD in local currency.

\subsection{Study Design and Procedure}
We conducted 20 standing sessions---including 10 in the \textit{watch-raised} pose and 10 in the \textit{arm-resting} pose, following the same protocol as in Study 1, which included 18 gestures with four repetitions for each session---to evaluate robustness to variations in posture, addressing: \textit{Can our system track hand poses when users are in different body postures?} For the collection of ground-truth data, we used an external webcam positioned below for the watch-raised pose and to the side for the arm-resting pose, with camera angles facing the palm to get the full finger joint ground truth.

\subsection{Results}
We collected a total of 288 minutes of data from 6 participants, including 144 minutes from the two postures: \textit{watch-raised} and \textit{arm-resting}. Following the first study, we evaluated the system performance on two subsets: (1) continuous fine-grained finger movement tracking using MPJPE and MPJAE metrics, and (2) gross wrist rotation using the MWAE metric. To evaluate WatchHand's robustness across varying body postures, we conducted a targeted analysis across three distinct conditions: (1) posture-independent model, (2) single-session fine-tuned model, and (3) nine-session fine-tuned model. The data collected for the two new postures---watch-raised and arm-resting---were analyzed separately throughout this evaluation. Given the small sample size, we focused on comparing the numerical values rather than formal statistical analysis.

We first evaluated generalization through a posture-independent model trained only on arm-raised sitting data (cross-session protocol, Section~\ref{sec:cross-session}) and tested on unseen postures without retraining (zero-shot evaluation), achieving \textcolor{blue}{9.96} mm MPJPE (SD 1.\textcolor{blue}{67}), \textcolor{blue}{7.72}$^\circ$ MPJAE (SD 1.\textcolor{blue}{24}), and 13.\textcolor{blue}{74}$^\circ$ MWAE (SD 6.\textcolor{blue}{11}). Next, with limited adaptation, fine-tuning on one new session \textcolor{blue}{(2 minutes of data for finger movements; 0.4 minutes of data for wrist rotation)} improved performance to \textcolor{blue}{7.65} mm MPJPE (SD 1.\textcolor{blue}{39}), 6.\textcolor{blue}{33}$^\circ$ MPJAE (SD \textcolor{blue}{1.43}), and \textcolor{blue}{8.06}$^\circ$ MWAE (SD 2.5\textcolor{blue}{7}). Lastly, more extensive adaptation with nine new sessions (\textcolor{blue}{18 minutes of data for finger movements; 3.6 minutes of data for wrist rotation, }matching the original training volume) further reduced errors to \textcolor{blue}{6}.58 mm MPJPE (SD \textcolor{blue}{1.12}), \textcolor{blue}{5.33}$^\circ$ MPJAE (SD \textcolor{blue}{1.05}), and \textcolor{blue}{5.83}$^\circ$ MWAE (SD \textcolor{blue}{1.72}). Overall, while the posture-independent model consistently demonstrated centimeter-level tracking accuracy, fine-tuning approaches revealed progressively improved performance as more data was used for training (see Figure~\ref{fig:study2_result}-B). This highlights the importance of data diversity and quantity for improving model generalization across body postures.
\section{\textcolor{blue}{Study 3: Robustness to Noises}}   
\label{sec:noise-study}
\textcolor{blue}{We further evaluated WatchHand under various noise conditions, reflecting common real-world usage where acoustic signals may be disturbed by external sounds and motion. This study was approved by the local IRB.}

\subsection{Participants}
\textcolor{blue}{We recruited 8 participants (5 male, mean age 23.88, SD 4.39) from a local university, matching the number of participants who used a single smartwatch in Study 1. Their mean hand length was 18.18 cm (SD 1.21), and their mean palm width was 8.04 cm (SD 0.62). Each session lasted two hours, and participants received 30 USD in local currency for compensation.}

\begin{figure*}[t]
\centering
   \includegraphics[width=\textwidth]{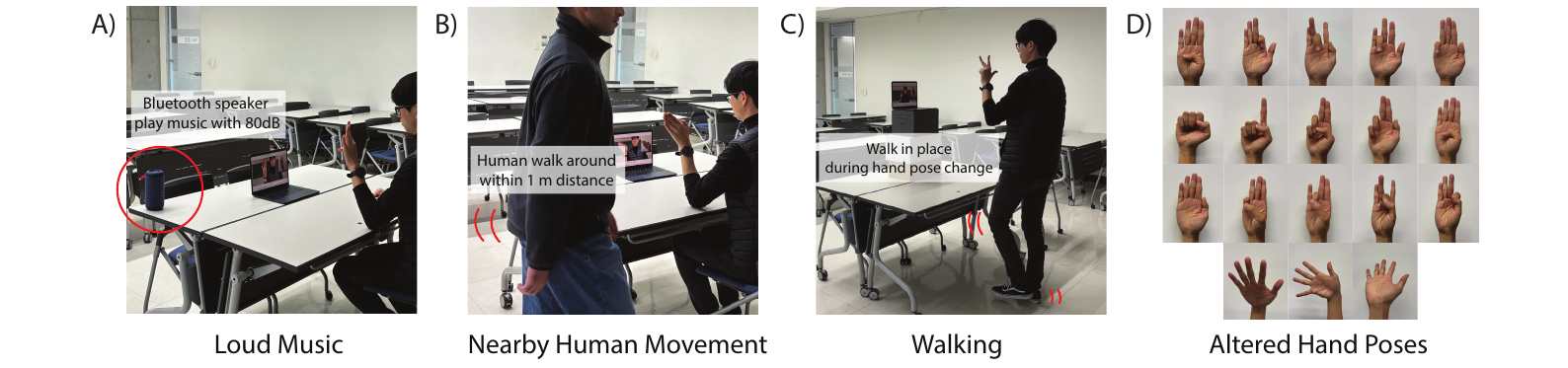}
   \hfil
\caption{\textcolor{blue}{Four noise and interference conditions used in Study 3: (A) high-volume audio played near the participant, (B) another person walking around the participant within 1 m, (C) participant walking in place while executing pose transitions, and (D) participants intentionally performing modified gestures (e.g., reduced or increased finger spacing) to simulate natural inconsistencies.}}
\Description{Figure 9 illustrates Four noise and interference conditions used in Study 3. (A) shows loud music condition, where high-volume audio played near the participant, (B) shows nearby human interference condition, where another person walking around the participant within 1 m, (C) illustrates participant walking in place while executing pose transitions, and (D) shows participants intentionally performing modified gestures (e.g., reduced or increased finger spacing) to simulate natural inconsistencies.}
\label{fig:noise_setup}
\end{figure*}

\subsection{Study Design and Procedures}
\textcolor{blue}{We generally followed the methodology outlined in Study 1. First, we collected 10 baseline sessions per participant in a seated posture using the same design as Study 1. To evaluate and compare robustness under noise conditions, we then collected 8 additional sessions per participant (2 for each of four noise scenarios), enabling both cross-condition and one-session fine-tuned evaluation, based on the promising fine-tuning results from Study 2. In total, each participant completed 18 sessions, following the same procedure outlined in Study 1, including consent, equipment setup, instructions, practice, main task, and questionnaire. The four noise/interference conditions were instructed as follows:}
\textcolor{blue}{(1) Loud music: high-volume audio (around 80 dB) was played through a Bluetooth speaker positioned one meter from the participant.}
\textcolor{blue}{(2) Nearby human movement: another person walked around the participant at around a one-meter distance.}
\textcolor{blue}{(3) Walking scenario: participants walked continuously while executing pose variations. They were instructed to walk in place at a natural walking pace to introduce motion-induced noise.}
\textcolor{blue}{(4) Altered hand poses: participants intentionally performed modified gestures to simulate natural inconsistencies. Specifically, they were instructed to eliminate finger spacing during hand-pose variations and to widen finger spacing during wrist-rotation poses, deviating from the poses instructed in previous studies.}

\begin{figure*}[t]
\centering
   \includegraphics[width=\textwidth]{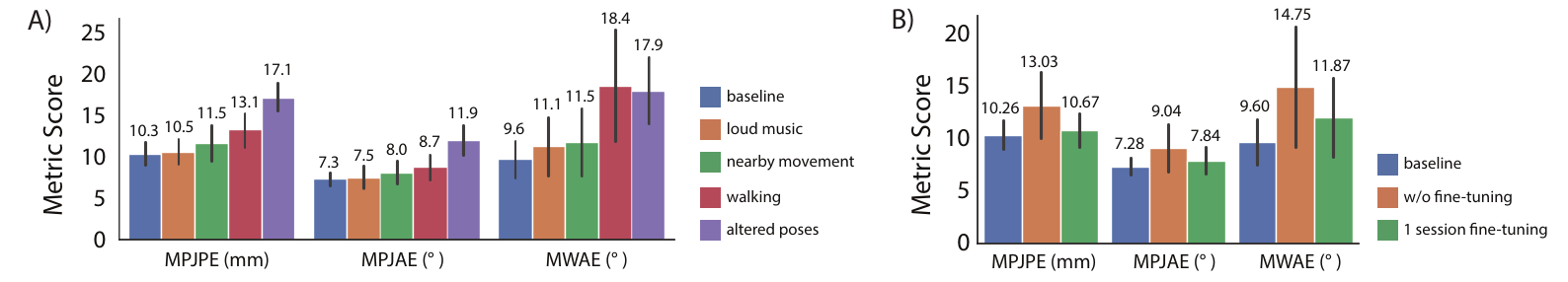}
   \hfil
\caption{\textcolor{blue}{Average hand pose tracking performances across all metrics, comparing baseline (no noise) to (A) four noisy conditions---loud music, nearby movement, walking, and altered poses, and (B) average performance across the noisy conditions with and without one-session fine-tuning.}}
\Description{Figure 10 shows average hand pose tracking performances across all metrics, (A) shows bar plot which comparing baseline (no noise) to four noisy conditions, including loud music, nearby movement, walking, and altered poses, and (B) shows comparing baseline to average performance across the noisy conditions with and without one-session fine-tuning.}
\label{fig:study4_result}
\end{figure*}

\subsection{Results}
\textcolor{blue}{We collected a total of 345.6 minutes of data from 8 participants, consisting of 192 minutes from the sitting condition and 38.4 minutes for each of the four noise/interference scenarios. In this study, we report three sets of results: 1. Baseline: training on 9 sessions and testing on the remaining session under the sitting condition; 2. Cross-condition: evaluating the same trained model directly on each of the four noise conditions; 3. One-session fine-tuning: adapting the model with one session (2 minutes of data for finger movements; 0.4 minutes of data for wrist rotation, following Study 2) from each noise condition and testing on the other session. All evaluations were conducted under a cross-session protocol to reflect realistic real-world usage.}

\textcolor{blue}{The baseline model achieved an average MPJPE of 10.26 mm (SD 1.37), MPJAE of 7.28° (SD 0.76), and MWAE of 9.6° (SD 2.11). Although collected under the same protocol as Study 1, this baseline performance exhibits a degradation, which we attribute to participant-dependent variability, implying that inter-participant articulation styles contribute to model performance. We then examined how this baseline model generalizes to the four noise scenarios (summarized in Figure~\ref{fig:study4_result}-A). Performance under loud music and nearby human movement remained statistically comparable to the baseline, with no significant differences observed across all metrics (p-values 0.18-0.76). This suggests that our system is robust to audible audio interference and that most nearby human motion occurring beyond the 21.42 cm sensing range is effectively filtered out during preprocessing. However, self-induced physical noise conditions led to significant degradations in the walking condition (p-values 0.003-0.028) and altered hand poses (all p < 0.001), indicating that physical interference within the sensing range (e.g., arm motion while walking or unseen variation in altered hand poses) can meaningfully disrupt the echo profiles. Finally, fine-tuning with a single session per noise condition substantially restored performance (See Figure~\ref{fig:study4_result}-B). Notably, the walking condition shows no significant difference from baseline across all metrics (p-values 0.13-0.54). These results demonstrate that WatchHand can accommodate real-world variability with minimal additional data, and that brief scenario-specific calibration can greatly improve accurate hand pose tracking.}

\section{\textcolor{blue}{Study 4}: Evaluation on Dynamic Hand Pose Variations} \label{sec:study3}

\begin{figure*}[t]
\centering
   \includegraphics[width=\textwidth]{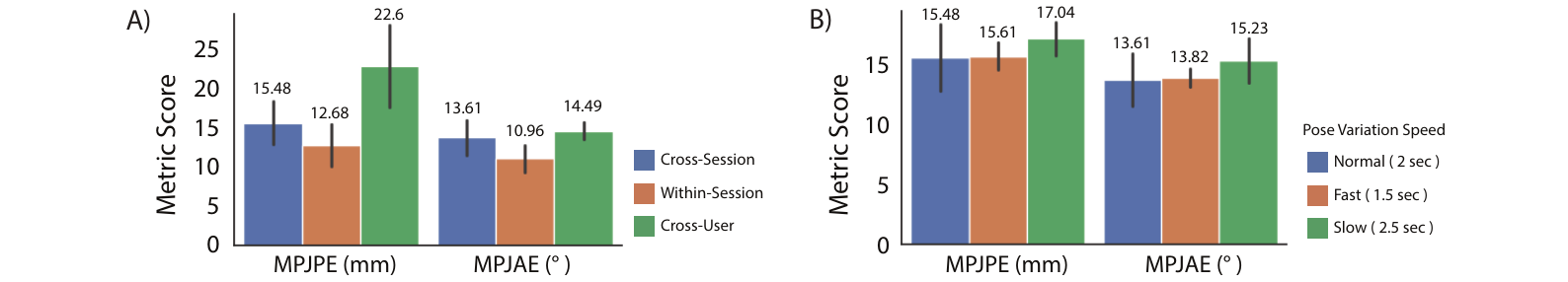}
   \hfil
\caption{Average hand pose tracking performance in Study \textcolor{blue}{4} across (A) evaluation protocols and (B) hand pose transition speeds.}
\Description{Figure 11 compares WatchHand’s hand pose tracking performance across evaluation protocols and pose variation speeds. (A) shows that cross-user evaluation yields the highest MPJPE (22.6 mm) and MPJAE (14.49°), reflecting inter-user variability, while within-session results are lowest (12.68 mm / 10.96°), confirming model reliability under stable conditions. (B) evaluates system robustness under different hand motion speeds. While performance degrades modestly for both faster (15.61 mm / 13.82°) and slower (17.04 mm / 15.23°) transitions, the system maintains bounded accuracy, indicating resilience to temporal variation.}
\label{fig:study3_result}
\end{figure*}

\textcolor{blue}{Our Studies 1-3} demonstrated promising results in continuous hand pose tracking using only the built-in microphone and speaker, following closely related protocols~\cite{Lee24_echoWrist}. However, the evaluated pose variations were constrained, as each of the 15 gestures (10 complex, 5 simple) returned to a neutral pose between transitions. To address this, our \textcolor{blue}{fourth} study captured direct pose-to-pose transitions---100 ordered transitions between the 10 complex gestures and 25 between the 5 simple gestures---for 125 total. This design exposes the model to richer, more realistic hand-motion dynamics. Building on the robustness to different watches and wearing hands established earlier, we selected a single representative device---Xiaomi Watch Pro 2 (chosen for its average tracking performance)---worn on the left hand. In addition, we evaluated the model’s robustness under different hand movement speeds, considering real-world variation. This study was approved by the local IRB.

\subsection{Participants}
We recruited 8 participants (4 male, all right-handed, mean age 23.5, SD 4.17) from a local university. Their mean hand length was 18.46 cm (SD 1.94), and their mean palm width was 7.99 cm (SD 0.94). Each session lasted approximately 90 minutes, and participants received a 25 USD voucher for their participation.

\subsection{Study Design}
We evaluated the same set of 15 hand gestures as in Study 1, consisting of 5 simple and 10 complex gestures. Wrist orientation gestures were excluded due to their limited pose-to-pose variation (only 6 combinations). Each session included diverse pose-to-pose transitions, including 4 repetitions of both simple and complex gestures, each lasting two seconds, presented in a randomized order, consistent with the Study 1 protocol. Participants completed 20 sessions at a normal transition speed (two seconds per pose change). To further assess robustness, each participant completed an additional session in both fast (1.5 s transition) and slow (2.5 s transition) conditions.

\subsection{Study Procedure}
This study followed a procedure similar to that of Study 1, including obtaining consent, donning equipment, delivering instructions, conducting practice trials, executing the main task, and completing a demographic questionnaire. Before beginning the experimental sessions, participants completed a practice trial. They then performed 20 normal-speed sessions, followed by one session each under fast and slow conditions. In the fast and slow conditions, participants were instructed to adjust their hand pose transition times to 1.5 s (fast) or 2.5 s (slow), following protocols in prior work~\cite{Lee24_echoWrist}. Considering practical and realistic variability, participants removed and re-wore the smartwatch between sessions.

\subsection{Results}
We collected a total of 352 minutes of data from 8 participants. This dataset includes 320 minutes of normal-speed sessions and additional test sessions at varying speeds and styles: 12 minutes of fast, 20 minutes of slow conditions. One participant (P28) exhibited a noticeable hand tremor when attempting to bend only the middle finger. To prevent muscle strain and ensure consistent data, we instructed him to stabilize the middle finger with his thumb (e.g., in ASL-8). We processed the captured data and implemented the acoustic model using the same procedures outlined in Study 1. Below, we report results under three evaluation protocols to provide a comprehensive view of WatchHand’s continuous 3D hand pose performances in pose-to-pose variation.

We first evaluated WatchHand under a \textit{cross-session} protocol, where the participants re-wore the device across different sessions, reflecting realistic everyday use. In this result, the model achieved an average MPJPE of 1\textcolor{blue}{5.48} mm (SD \textcolor{blue}{2.75}) and an MPJAE of 1\textcolor{blue}{3.61}$^\circ$ (SD 2.\textcolor{blue}{18}). Next, we evaluated WatchHand under a \textit{within-session} protocol---as an upper-bound performance benchmark---since device placement and environmental factors remain constant. In this result, the model achieved an average MPJPE of 12.\textcolor{blue}{68} mm (SD 2.\textcolor{blue}{63}) and an MPJAE of 1\textcolor{blue}{0.96}$^\circ$ (SD 1.\textcolor{blue}{72}). Finally, we performed a \textit{cross-user} evaluation, where models were trained on data from all other participants and tested on unseen participants, providing the system’s generalizability to new users. In this result, the model achieved an average MPJPE of 2\textcolor{blue}{2.60} mm (SD 5.1\textcolor{blue}{8}) and an MPJAE of 14.\textcolor{blue}{49}$^\circ$ (SD \textcolor{blue}{1.12}). Overall, these results show consistent tendencies with Study 1 across protocols: strongest performance in within-session settings, moderate error increases in cross-session evaluation, and the largest drop in cross-user testing (see Figure~\ref{fig:study3_result}-A)---patterns consistent with prior work~\cite{Devrio22DiscoBand, Hu20FingerTrak, Kyu24EITPose}.

\begin{figure*}[t!]
\centering
   \includegraphics[width=\textwidth]{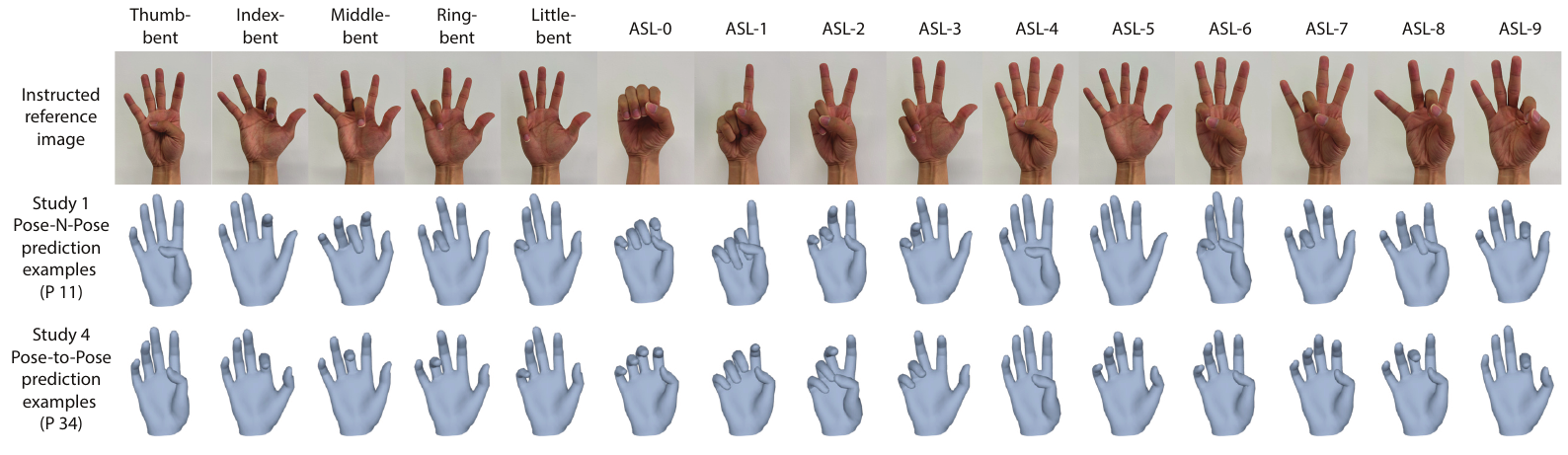}
   \hfil
\caption{\textcolor{blue}{Hand poses instructed in the studies (top row) and example predictions (visualized with MANO~\cite{Romero17MANO}) from the pose-neutral-pose cross-session model in Study 1 (middle row) and pose-to-pose cross-session model in Study 4 (bottom row).}}
\Description{Figure 12 presents the fifteen hand poses used in the study for fine-grained finger movement tracking. The top row shows the instructed reference images for each pose, including finger-specific bends and ASL number gestures (ASL-0 to ASL-9). The middle and bottom row displays the corresponding hand pose predictions from the WatchHand deep-learning model, demonstrating close alignment with the reference image. They are visualized with MANO from the pose-neutral-pose cross-session model in Study 1 at middle row, and pose-to-pose cross-session model in Study 4 at bottom row.}
\label{fig:qualitative_vis}
\end{figure*}

We then analyzed the robustness of WatchHand under different hand pose variation speeds. We used a cross-session model for this testing. For different speeds, the model achieved an average MPJPE of 1\textcolor{blue}{5.61} mm (SD \textcolor{blue}{1.13}) for the fast condition, 1\textcolor{blue}{7.04} mm (SD 1.\textcolor{blue}{35}) for the slow condition. The corresponding MPJAE results were 13.\textcolor{blue}{8}2° (SD \textcolor{blue}{0.73}) and 1\textcolor{blue}{5.23}° (SD 1.\textcolor{blue}{82}), respectively. These results show that WatchHand maintains tracking during fast pose changes, with modest performance drops under slower transitions (see Figure~\ref{fig:study3_result}-B), highlighting both the system’s robustness and the need for broader training data to capture diverse real-world variability. As expected, the more dynamic pose-to-pose transitions in Study \textcolor{blue}{4} introduced additional variability compared to the pose-neutral-pose protocol in Studies 1\textcolor{blue}{-3}. Unlike Study 1, where each gesture transition was anchored at a neutral open-hand pose, Study \textcolor{blue}{4} involved direct transitions between arbitrary gesture pairs (125 possible combinations), which led to noisier intermediate states and a broader range of temporal dynamics. Overall, these results confirm that pose-to-pose transitions are indeed more challenging, but WatchHand still achieves sub-centimeter accuracy in within-session tests and competitive cross-session/user performances. 

\textcolor{blue}{We further examined WatchHand’s behavior by qualitatively inspecting prediction frames. Figure~\ref{fig:qualitative_vis} illustrates example reconstructions from a single participant in the pose-neutral-pose cross-session model (Study 1) and the pose-to-pose cross-session model (Study 4), including failure cases, along with the corresponding instruction images. The participants (P11 and P34) were chosen because their overall tracking performance was close to the average score for each study. Overall, our system is particularly effective at predicting thumb movement, likely because the thumb is closer to the smartwatch and follows a distinct bending direction compared to the other fingers, and appears more frequently during training (both in thumb-bent and ASL-4 poses). However, several key error patterns were observed. First, neighboring-finger effects: the model occasionally activated adjacent fingers incorrectly---for example, predicting motion in both the index and middle fingers when only the index should move, likely reflecting biomechanical coupling between fingers and the limited separability of their acoustic signatures, and is especially common around the middle finger, where inter-finger spacing is small~\cite{jimenez2019relationships}. Second, ergonomic constraints: since many participants could not bend the little or ring finger in isolation, the model sometimes reflected these real-world limitations, leading to misrepresentation in those poses. Third, anomalies: we observed occasional random errors (e.g., a single-finger bend predicted during a complex gesture), which we suspect are caused by transient noise, such as subtle posture adjustments or arm movements. These anomalies appear more frequently in the pose-to-pose model, where dynamic transitions are richer, and training data per transition is sparser compared to Study 1. In contrast, in the pose-neutral-pose model, the default open-palm pose (ASL-5) is sometimes over-predicted when the reflected signals are weak (e.g., for isolated little-finger bends). Overall, these qualitative observations are consistent with inter-participant differences in behavior and finger articulation.} 

\section{Discussion and Limitation}
Our studies confirm the encouraging performance of WatchHand for inferring continuous 3D hand poses across three common COTS smartwatches. Because it operates entirely with built-in hardware and does not require bespoke sensors, WatchHand has strong potential to immediately expand the interaction capabilities of COTS smartwatches. The results also suggest its applicability across diverse scenarios, including more complex hand poses, different smartwatch models, watch-wearing hands, body postures, \textcolor{blue}{and noise conditions}. In this section, we further discuss the opportunities and challenges of deploying WatchHand more broadly across diverse hand poses and use contexts to support everyday use in the future.

\subsection{Generalizing to Diverse \textcolor{blue}{Hand Poses, Users, and Contexts}}

\begin{figure*}[t]
\centering
   \includegraphics[width=\textwidth]{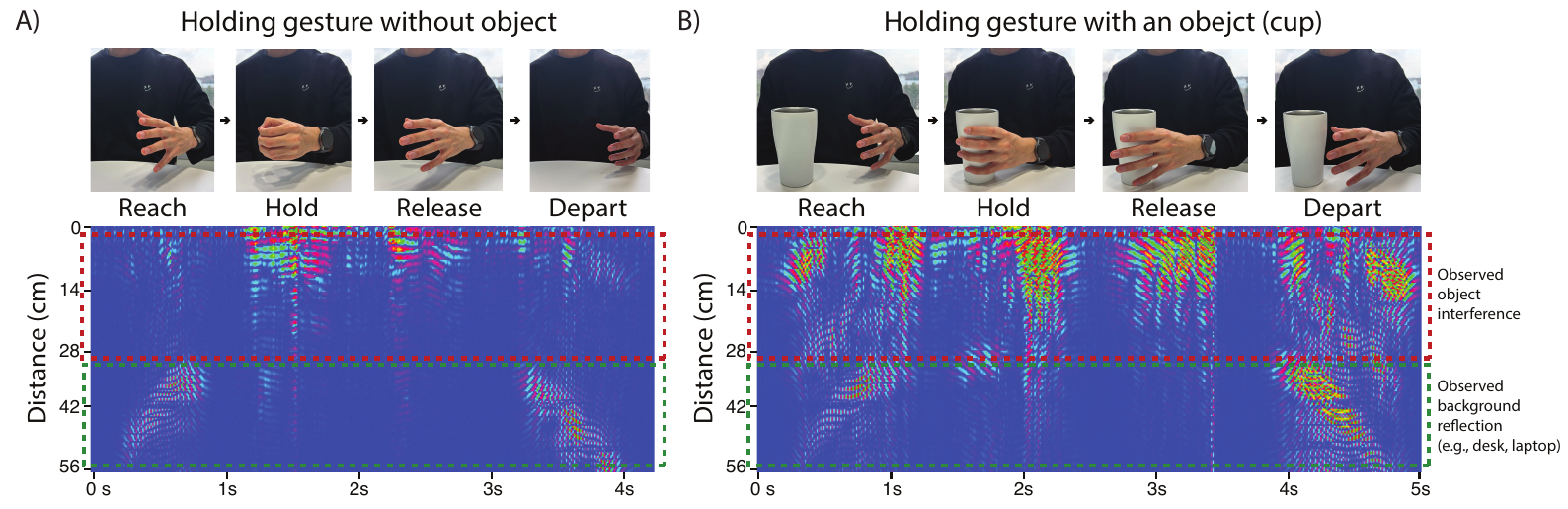}
   \hfil
\caption{\textcolor{blue}{Impact of hand–object interaction on acoustic hand pose tracking. (A) Echo profiles of a holding gesture performed without an object and (B) with a cup. Each sequence consists of four phases: Reach, Hold, Release, and Depart. Compared to the object-free condition, performing the gesture while holding an object produces noticeable acoustic interference near the smartwatches.}}
\Description{

}
\label{fig:object_interference}
\end{figure*}

\subsubsection{Dynamic Hand Pose Variation}
We evaluated WatchHand under two hand pose variation protocols: pose-neutral-pose (returning to a neutral hand pose between pose variations) in Studies \textcolor{blue}{1-3} and pose-to-pose (direct transitions between poses) in Study \textcolor{blue}{4}. The latter introduces richer intermediate states and greater temporal variability, which, as expected, led to a modest accuracy drop (96.7\% increase in MPJPE in cross-session evaluation) relative to pose-neutral-pose. However, this drop must be understood in the context of the greatly increased complexity of the task: the number of possible pose transitions increased by 8.3 times (from 15 to 125). Despite this more challenging setting, WatchHand outperforms prior pose-to-pose systems (e.g., DiscoBand~\cite{Devrio22DiscoBand} (17.87 mm) and EITPose~\cite{Kyu24EITPose} (17.81 mm)) in cross-session evaluation, demonstrating its robustness and the feasibility of scaling continuous hand pose tracking to real-world smartwatch use.

\subsubsection{Unseen Hand Poses}
While our study includes a diverse set of hand poses, comprehensively covering the entire space of possible hand configurations remains challenging. \textcolor{blue}{In Study 3, for example, altered hand poses (e.g., reducing or enlarging gaps between fingers) led to a 66.19\% increase in MPJPE in cross-condition testing, yet a single-session (2 minutes of data) fine-tuning reduced this to only a 5.3\% drop relative to the baseline.} In Study \textcolor{blue}{4}, we observed only a modest drop in tracking performance under different hand pose variation speeds---mean MPJPE increased by 2.6\% in the fast conditions and 11.5\% in the slow conditions---reflecting the added ambiguity of intermediate states and trajectory diversity. Importantly, these effects were bounded rather than catastrophic, suggesting the sensing pipeline remains informative even when encountering out-of-distribution poses. These results demonstrate strong generalization to unseen poses, an important step toward real-world usability with larger, more diverse datasets.

\subsubsection{Unseen Users}
Cross-user evaluation yielded encouraging results, with MPJPE ranging from 15 to 23 mm across both pose-neutral-pose and pose-to-pose protocols. Performance degradation was primarily due to inter-individual differences in hand size and articulation, which introduced greater variability in pose execution. \textcolor{blue}{This pattern is consistent with prior wrist-worn hand-pose tracking systems, where cross-user performance is substantially weaker than within-session~\cite{Devrio22DiscoBand, Kyu24EITPose} or cross-session results~\cite{Lee24_echoWrist, Hu20FingerTrak}.} These findings underscore a key challenge of cross-user generalization: while the system adapts well within a user, accommodating inter-individual differences remains challenging. Addressing this will require larger and more heterogeneous datasets spanning a wider range of users and hand characteristics, \textcolor{blue}{and meta-learning–style adaptation could further improve cross-user robustness.}

\subsubsection{Robustness to Noise}
\textcolor{blue}{Our noise study (Study 3) suggests that the primary vulnerability of smartwatch-based acoustic hand tracking is not ambient sound or distant movement, but physical changes near the sensing region. Loud music and nearby human movement had a negligible impact, indicating that WatchHand, operating in the inaudible band and with a 21.42 cm sensing range, is effectively decoupled from most everyday audio activity and background motion. In contrast, mobile scenarios (e.g., walking) altered the propagation paths around the wrist and hand, leading to measurable degradations and confirming that nearby physical changes around the wrist and hand disrupt the echo signals. Nevertheless, a single short fine-tuning session per condition largely restored performance, showing that WatchHand can adapt to real-world noise and motion scenarios with minimal calibration, reinforcing its practicality for everyday smartwatch use.}

\subsubsection{\textcolor{blue}{Holding Object Interferences}}

\textcolor{blue}{While WatchHand reliably tracks 3D hand poses in free-hand scenarios, its performance will likely be substantially impacted when users hold objects. An exploratory analysis (see Fig~\ref{fig:object_interference}) supports this observation: when participants performed the same reach–hold–release–depart sequence~\cite{grasphUI24Adwait} with and without a cup, the object introduced localized interference in the echo profiles. These reflections stemmed from object surfaces rather than the hand, masking critical pose-dependent patterns and confirming that object interaction is a meaningful disruptor to single-channel acoustic sensing. Quantifying the resulting degradation is challenging, as object manipulation fundamentally changes hand geometry and invalidates the original pose protocol, making direct comparison with free-hand conditions infeasible. However, this limitation is not unique to our approach. Commercial AR/VR systems typically suspend hand tracking under occlusion rather than produce unreliable poses~\cite{Meta}. A smartwatch-based acoustic system could adopt a similar strategy—detecting object presence and temporarily disabling pose tracking. This limitation highlights an exciting direction for future research. Integrating object-awareness---such as recognizing grasp states or supporting micro-gestures during holding, as explored in bespoke systems~\cite{lee2025grab}, could enable WatchHand to transition seamlessly between free-hand and object-mediated interactions in real-world use.}

\subsubsection{Strategies for \textcolor{red}{Improving and} Enriching Datasets}
Despite our extensive evaluations, the dataset size remains a key limitation due to resource constraints. As prior work~\cite{Kim22EtherPose, Devrio22DiscoBand, Lee24_echoWrist, Kyu24EITPose, Hu20FingerTrak} has noted, larger-scale datasets are essential for training more sophisticated models to handle unseen users, poses, \textcolor{blue}{and contexts}. Given that our system is implemented with COTS smartwatches and optical tracking systems, we can leverage this accessibility to implement a distributed data-collection approach, potentially through an opt-in framework that allows users to contribute anonymized hand-pose data during everyday use. Beyond collecting additional real-world data, developing a physics-based or data-driven acoustic simulator could generate synthetic echo profiles for \textcolor{blue}{underrepresented hand poses, helping to fill gaps in pose coverage. \textcolor{red}{Furthermore, lab studies using a research-grade optical tracking system (e.g., Optitrack or similar) could contribute more accurate and reliable baseline data.} In parallel, self-supervised learning could enable the model to learn meaningful representations from unlabeled acoustic data, reducing reliance on dense ground-truth annotations. We further acknowledge that unsupervised domain adaptation methods, which encourage invariance across users, sessions, or poses, are a promising direction to narrow the remaining generalization gap, especially in cross-user settings.} Together, these strategies would strengthen generalization and pave the way toward broader real-world deployment of WatchHand. 

\subsection{\textcolor{blue}{Toward Real-World Implementation}}

\subsubsection{\textcolor{blue}{Bespoke Hardware vs Off-the-shelf Smartwatches}}

\textcolor{blue}{Recent acoustic hand-tracking systems such as EchoWrist~\cite{Lee24_echoWrist} achieve high-quality pose estimation by using two speaker–microphone pairs---one placed on the watch body and another embedded in the wrist strap to directly face the palm. Although this configuration produces stronger acoustic reflections and delivers improved tracking performance compared to WatchHand (4.81 vs 7.87 mm MPJPE in cross-session evaluation), it relies on non-commercial hardware modifications. Embedding electronics into straps \textcolor{red}{is uncommon in} today’s smartwatches, as straps are intentionally user-replaceable, vary across brands, and are not \textcolor{red}{currently designed or optimized} to host powered sensing components. In contrast, WatchHand demonstrates that continuous 3D hand pose tracking can be achieved using only a single speaker and microphone already embedded in commercial smartwatches. This distinction is critical. Bespoke systems illustrate what enhanced sensing could achieve if hardware were redesigned; WatchHand shows what is possible today, on devices that already ship at scale. By enabling fine-grained 3D hand pose tracking on millions of existing smartwatches, WatchHand represents the first practical pathway for transitioning continuous hand tracking from bespoke academic prototypes to real-world smartwatch platforms.}

\subsubsection{Hardware Configuration and Watch-wearing Hands} 
WatchHand demonstrates robustness to variations in hardware platforms and watch orientations. Our results show no significant impact from differences in hardware configurations across manufacturers or from flipped orientations due to the watch-wearing hand (left vs. right). This highlights WatchHand's potential for universal applicability. \textcolor{blue}{In our main experiment, we train separate models for each watch–hand configuration, so each model learns a stable mapping for its specific setup. To further examine generalizability, we trained a general model with training data aggregated across all smartwatch models and watch-wearing hands in cross-session testing.} The results are promising, demonstrating a comparable performance to the per-configuration cross-session models, with an average MPJPE of 7.\textcolor{blue}{88} mm (SD 1.4\textcolor{blue}{9}), MPJAE of 6.22$^\circ$ (SD 1.4\textcolor{blue}{5}), and MWAE of \textcolor{blue}{7.68}$^\circ$ (SD \textcolor{blue}{1.35}). \textcolor{blue}{Since all tested devices share a similar acoustic layout (speaker and microphone on opposite edges), these results suggest that a single model can generalize across brands and orientations, supporting scalable deployment on COTS smartwatches.}

\subsubsection{System Deployment}
In this study, as in many prior works~\cite{Lee24_echoWrist, Yu24_ringAPose, Kim22EtherPose, Devrio22DiscoBand, Kyu24EITPose, Liu21WRHand, Hu20FingerTrak}, our system initially streamed the raw sensor data from the smartwatch to external servers (e.g., laptop or smartphone), with signal processing and model inference performed on a remote server. However, to be deployed in real-world settings, the system needs to run entirely on-device without relying on external infrastructure. To demonstrate the feasibility of this, we deployed the entire pipeline model---including signal processing and model inference---on commercial smartwatches. The model (26.7 MB) executes fully on-device (tested on Samsung Galaxy Watch 7), enabling real-time, continuous hand pose tracking without transmitting data to external systems. The average latency per prediction is 0.115s (SD 0.006), comprising 0.035s (SD 0.004) for preprocessing and 0.08s (SD 0.005) for inference, supporting responsive interaction in practical use.

\subsection{Practical Consideration}

\subsubsection{Power Consumption}
Running continuous playback and recording for active acoustic sensing on a smartwatch introduces non-trivial power demands. To estimate this cost, we monitored the battery usage over a 3-minute session---including key operations such as recording, playback, and data transmission, along with ambient power consumption from screen lighting and background processes. The Galaxy Watch 7 consumed approximately 1\% of its battery (4.25 mAh of 425 mAh), the Xiaomi Watch 2 Pro about 0.87\% (4.31 mAh of 495 mAh), and the Pixel Watch 3 about 1.52\% (6.36 mAh of 420 mAh). These differences in power consumption may be attributed to variations in hardware configurations and sensor characteristics---for instance, the Pixel Watch 3 exhibited a higher speaker output level, which may contribute to its increased energy use. It is important to note that these numbers reflect the entire system on the COTS watch (e.g., OS, display, and other functions), and our system is not yet optimized for this sensing task, as typical wearable microphones and speakers are reported to have power signatures~\cite{Lee24_echoWrist, Yu24_ringAPose} under 1 mW. This indicates that further optimization could significantly extend the usage time for a smartwatch-based acoustic hand pose tracking system. \textcolor{blue}{In practice, to conserve power, hand pose tracking with acoustic sensors could be activated only when low-power IMU-based gestures~\cite{ApplewatchAssitiveTouch, samsungMoreThan} are detected.}

\subsubsection{Privacy}
Operating in the 18-21 kHz frequency range offers WatchHand privacy advantages, as this ultrasonic range lies significantly above the typical spectrum of human speech, which spans approximately 0.1 to 8 kHz. Our system implements frequency filtering to isolate only signals within our designated operating range, effectively eliminating speech-related audio content from processing. Although the current WatchHand system initially records broadband audio, including speech frequencies, our processing pipeline immediately discards this lower-frequency information. In future implementations, when the entire system can be deployed directly on the smartwatch without requiring external processing, privacy protection can be further enhanced through on-device computation, ensuring that potentially sensitive audio data never leaves the user's personal device.

\subsubsection{Potential Applications}
Building on the strong and robust hand pose tracking performance results we report, we suggest potential practical applications of WatchHand in daily wearable contexts. 
One key application area is gesture-based interaction. WatchHand allows users to interact with devices through natural hand movements without requiring physical touch, which is particularly valuable for hands-busy or touch-free scenarios, such as controlling music playback while cooking, navigating slides during a presentation, or issuing commands \cite{devrio2023smartposer, sharma2023sparseimu, arakawa2024prism}. Continuous tracking also enables more expressive gestures beyond discrete taps or swipes, allowing for nuanced and expressive controls such as pinch-to-zoom, drag, or swipe variations based on finger articulation \cite{yu2017tap}. 
In addition, WatchHand can support assistive technologies, particularly for users with limited mobility or speech. By detecting subtle and continuous finger movements, the system could facilitate silent communication (e.g., fingerspelling~\cite{Pugeault11Spelling, lim2025spellring}) and customizable input for accessibility settings. Finally, WatchHand has the potential to extend cross-device interaction. When paired with smartphones, tablets, or smart glasses, smartwatches running WatchHand could serve as lightweight and low-cost, wrist-mounted input devices for spatial interfaces \cite{salter2024emg2pose}. These applications outline how continuous hand pose tracking---achieved solely by the sensors within existing COTS smartwatches---can empower more natural, expressive, and context-aware interaction in a range of real-world scenarios.
\section{Conclusion}
We presented WatchHand, the first continuous 3D hand pose tracking system running entirely on commercial smartwatches using only the built-in speaker and microphone. Leveraging active acoustic sensing, WatchHand reliably tracks finger movements and wrist rotations without requiring external hardware. Our evaluations across different smartwatch models, watch-wearing hands, body postures, \textcolor{blue}{noise conditions,} and hand pose variation protocols confirm the system’s robustness, adaptability, and generalizability. By utilizing existing smartwatch hardware, WatchHand offers a practical, privacy-preserving, and scalable solution that is capable of immediate deployment. This breakthrough substantially lowers the barrier to smartwatch-based hand tracking and unlocks new opportunities for always-available interactions on millions of devices already in use today.


\begin{acks}
We thank all the participants who took part in our user studies. This work was supported by the National Science Foundation under Grant No.2239569 (NSF CAREER Award) and the IITP(Institute of Information \& Communications Technology Planning \& Evaluation)-ITRC(Information Technology Research Center) grant funded by the Korea government(Ministry of Science and ICT)(IITP-2026-RS-2024-00436398).
\end{acks}

\bibliographystyle{ACM-Reference-Format}
\bibliography{references}

\end{document}